%% file: paper.tex
\documentclass[aip,jcp,reprint,twoside,groupedaddress,citeautoscript,10pt]{revtex4-1}
\let\cedilla\c

\input{packages}
\input{definitions}

\begin{document}

\title{Non-additive dissipation in open quantum networks out of equilibrium}

\author{Mark T. Mitchison}
\email[]{markTmitchison@gmail.com}
\author{Martin B. Plenio}
\affiliation{Institut f\"ur Theoretische Physik, Albert-Einstein Allee 11, Universit\"at Ulm, D-89069 Ulm, Germany}

\pagestyle{fancy}
\fancyhead[RE]{\textsc{Non-additive dissipation in open quantum networks}}
\fancyhead[LO]{\textsc{Mitchison \& Plenio}}
\fancyhead[RO]{\thepage}
\fancyhead[LE]{\thepage}
\fancyfoot{}

\begin{abstract}
\small We theoretically study a simple non-equilibrium quantum network whose dynamics can be expressed and exactly solved in terms of a time-local master equation. Specifically, we consider a pair of coupled fermionic modes, each one locally exchanging energy and particles with an independent, macroscopic thermal reservoir. We show that the generator of the asymptotic master equation is not additive, i.e.\ it cannot be expressed as a sum of contributions describing the action of each reservoir alone. Instead, we identify an additional interference term that generates coherences in the energy eigenbasis, associated with the current of conserved particles flowing in the steady state. Notably, non-additivity arises even for wide-band reservoirs coupled arbitrarily weakly to the system. Our results shed light on the non-trivial interplay between multiple thermal noise sources in modular open quantum systems.
\end{abstract}

\maketitle


\section{Introduction}

An improved understanding of the dynamics of open quantum networks is desirable for research fields including quantum thermodynamics~\cite{GemmerMichelMahler, Kosloff2014arpc,Goold2016jpa,Benenti2017}, quantum biology~\cite{Huelga2013cp}, mesoscopic electronics~\cite{Datta} and the theory of non-equilibrium phase transitions~\cite{Mitra2006prl,Prosen2008prl,ProsenIlievski2011prl,Znidaric2011pre,MendozaArenas2013prb,Kirton2013prl}. The Lindblad master equation~\cite{Lindblad1976cmp,Gorini1976jmp} is a popular and powerful tool for modelling such systems, which approximates the long-time dynamics under the assumption of weak coupling to memoryless environments, i.e.\ the Born-Markov approximation (BMA). Additional, uncontrolled approximations are typically required in order to obtain a completely positive evolution, such as the secular~\cite{BreuerPetruccione}, singular-coupling-limit~\cite{RivasHuelga} or wide-band-limit~\cite{Gurvitz1996prb,StefanucciVanLeeuwen} approximations. However, in \textit{composite} open systems, distinct approximations may lead to different and sometimes drastically incorrect predictions. 

In particular, standard master equations derived under the BMA describe dissipation in terms of either local processes on a small number of sites or global transitions between energy eigenstates of the entire network~\cite{Rivas2010njp}. Unfortunately, the local approach appears to violate thermodynamic laws~\cite{Levy2014epl,Stockburger2017fp} for certain kinds of time-independent system-bath interactions~\cite{Barra2015,Strasberg2017}, while the global approach fails to capture the coherences necessary to properly describe the non-equilibrium steady state (NESS)~\cite{Wichterich2007pre}. This poses a particular problem for the ongoing study of quantum thermal machines, where the multifaceted role played by coherence---acting variously as a performance-enhancing resource~\cite{Scully2003sci,Brunner2014pre,Uzdin2015prx,Mitchison2015njp,Raam2016prapp,Korzekwa2016njp,Maslennikov2017,Watanabe2017prl}, a useful output~\cite{Brask2015njp,Tavakoli2017,Misra2016pra,Manzano2017} or an unavoidable hindrance~\cite{Correa2013pre,Brandner2017}---remains incompletely understood.

Here, we aim to elucidate these issues by studying an exactly solvable model of a non-equilibrium quantum network. More precisely, we derive a time-local master equation describing a pair of coupled, localised fermionic modes. Each mode exchanges particles and energy with a macroscopic reservoir represented by a semi-infinite, uniform tight-binding chain, such that the total Hamiltonian is quadratic in fermionic ladder operators. This represents a prototypical quantum thermal machine, which could be realised by electrons flowing through a serial double quantum dot~\cite{Thierschmann2013njp} or cold fermionic atoms confined by an optical lattice~\cite{Brantut2012sci,Brantut2013sci}. 

Note that other authors have already derived and extensively studied exact time-local master equations describing general quadratic Fermi systems~\cite{Tu2008prb,Jin2010njp,Yang2013pra,Ribeiro2015prb}. In contrast to these previous approaches, the simplicity and symmetry of our specific set-up enables compact analytical solutions to be obtained using elementary methods and without needing the Born-Markov, secular, singular-coupling-limit or wide-band-limit approximations. Instead, we use an alternative, perturbative approximation scheme, valid when the system-environment coupling is much smaller than the bandwidth of the reservoir vacuum noise, and explicitly confirm its accuracy using the exact solution. This simplification allows us to clearly identify and distinguish the different physical processes governing the dynamics.

Our analysis challenges a central tenet of standard open-systems theory: namely, that two initially uncorrelated environments should give rise to two independent, additive contributions to the master equation in the weak-coupling limit (see, for example, Ref.~\onlinecite{Schaller}). Instead, we distinguish three contributions to the generator of asymptotic time evolution. Two of these describe the individual thermalising effect of each reservoir, while the third is an interference term arising from the combined action of both reservoirs whenever they are initially out of equilibrium with each other. This interference gives rise to coherence in the energy eigenbasis of the network, a feature which reflects the conserved fermion current flowing in the NESS. 

These findings connect several previous theoretical studies on composite open quantum systems. In particular, a significant body of research has sought to assess the validity of approximate, additive master equations by comparing them to each other or to exact numerical results~\cite{Wichterich2007pre,Scala2007jpa,Rivas2010njp,Levy2014epl,Stockburger2017fp,Purkayastha2016pra,Trushechkin2016epl,Eastham2016pra,Decordi2017oc,Hofer2017,Gonzalez2017}. Viewed broadly, these investigations indicate that the local and global Lindblad equations are only accurate in limited, complementary parameter regimes, and only for certain observables, even though the requisite conditions for the BMA might hold for each bath individually. On the other hand, a few recent papers have shown that the additivity assumption fails in the presence of multiple strong or structured noise sources~\cite{Chan2014pra,Giusteri2017pre,Brask2017arX}. 

We extend these results by demonstrating that interference between different thermal baths out of equilibrium gives rise to non-additive dynamics, even when the open system couples arbitrarily weakly to spectrally unstructured reservoirs. The interference contribution is not in Lindblad form and thus underlies the occurrence of asymptotic non-Markovianity found by Ribeiro \textit{et al.}~\cite{Ribeiro2015prb}, where the long-time dynamics cannot be described by a Markovian master equation even when the open system has lost all memory of its initial state~\cite{Hall2014pra,Petrillo2016arX,Ferialdi2017pra}. Moreover, we show that our non-additive master equation interpolates between the local and global Lindblad equations and recovers them in different limits. Since additivity is a necessary consequence of the Born-Markov approximation~\cite{Brask2017arX}, this implies the failure of the BMA for describing steady-state transport away from these limits. Nevertheless, we find that certain observables---such as the steady-state currents flowing into the baths---are accurately predicted by an additive master equation across a relatively broad range of parameters. Our work thus helps to clarify the validity of existing Lindblad models, while providing a reference point for future research aimed at moving systematically beyond the standard approximations. 
 
The plan of the paper is as follows. Section~\ref{sec_prelim} introduces some preliminary concepts and overviews our main results. The microscopic model that forms the core of this work is defined and exactly solved in Section~\ref{sec_model}. We derive the exact master equation describing the system and develop a weak-coupling approximation scheme in Section~\ref{sec_ME}. Non-additivity of the asymptotic dynamics is explored in Section~\ref{sec_nonadditive}. We discuss our results and conclude in Section~\ref{sec_conclusion}. 


\section{Preliminaries}
\label{sec_prelim}

\begin{figure}
\includegraphics[width=0.5\linewidth]{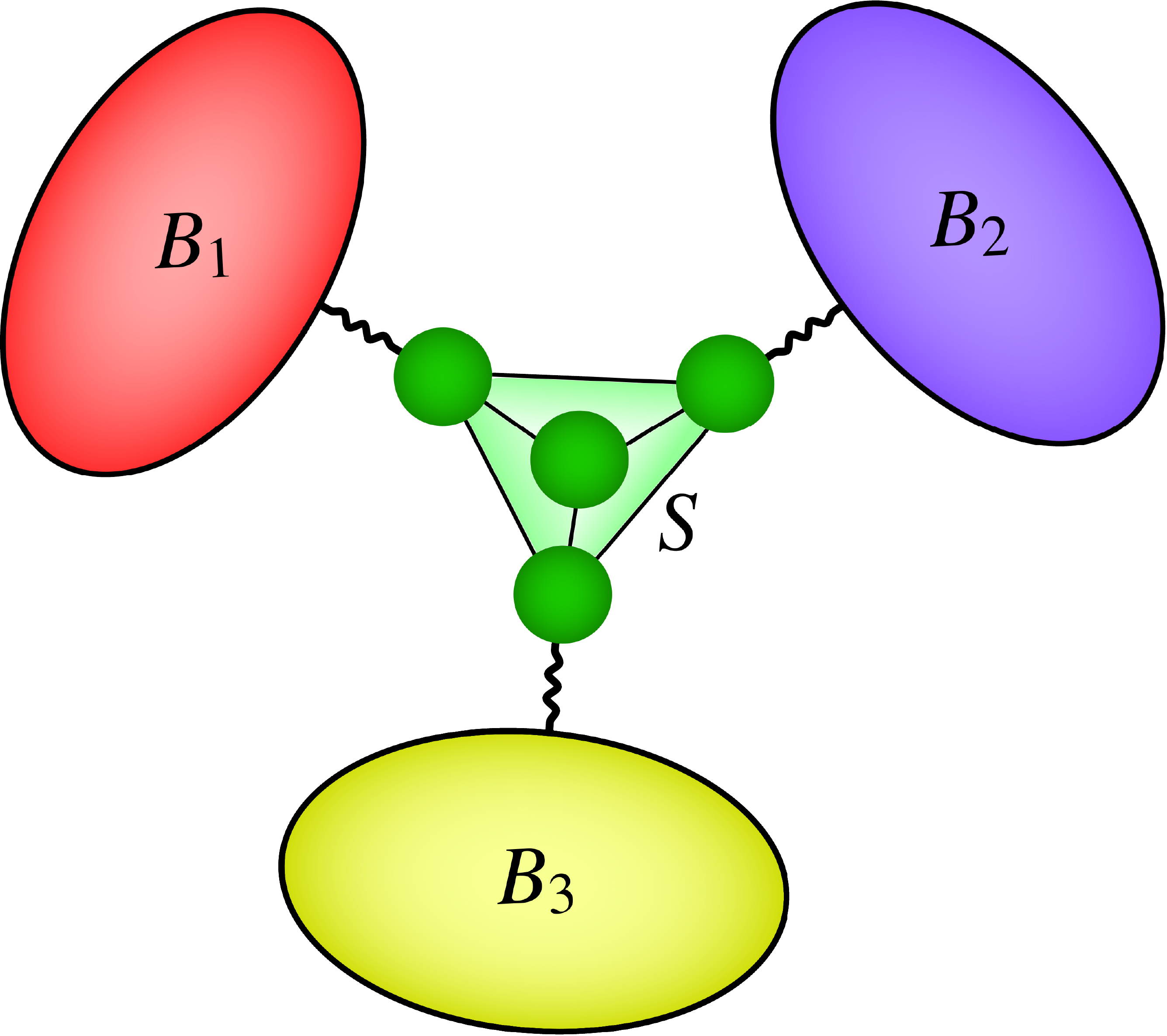}
\caption{An open quantum system $S\!$ comprising a network of interacting modules connected to different thermal reservoirs $B_{\alpha}$.\label{fig_openNetwork}}
\end{figure}

To motivate the problem at hand, consider an open quantum system $S\!$ comprising a network of coupled sites. This network is coupled weakly to multiple thermal reservoirs which may exchange particles and energy with $S\!$, as illustrated in Fig.~\ref{fig_openNetwork}. Each reservoir $B_\alpha$ is characterised by a temperature $T_\alpha = 1/\beta_\alpha$ and chemical potential $\mu_\alpha$ (we work in units where $k_B = 1$ and $\hbar=1$). Let $\H_{S\!}$ and $\hat{N}_{S\!}$ respectively be the Hamiltonian and particle number operators on $S\!$, where $[\H_{S\!},\hat{N}_{S\!}]=0$ unless $\mu_\alpha = 0$. We assume for simplicity that $\H_{S\!}$ has a non-degenerate spectrum. The quantum state of the network at time $t$ is denoted $\r_{S\!}(t)$. We suppose that initially each of the baths is not correlated with the others nor with the open system, and that $S\!$ asymptotically approaches a unique stationary state $\r_{S\!}^\infty = \lim_{t\to\infty}\r_{S\!}(t)$. We now summarise the behaviour that we expect in general.

One fundamental property of a thermal reservoir is that a small system weakly coupled to it should eventually equilibrate to the same temperature and chemical potential. In addition, the composition of several independent thermal reservoirs in equilibrium itself constitutes a thermal reservoir. Hence, \textit{for equal reservoir temperatures and chemical potentials the system should equilibrate}, i.e.\
\begin{equation}
\label{rhoInfEquilibrium}
\r_{S\!}^\infty = \frac{\ee^{-\beta \left (\hat{H}_{S\!}-\mu \hat{N}_{S\!}\right )}}{\ZZ\!\left (\beta,\H_{S\!},\mu,\hat{N}_{S\!}\right )} \quad {\rm if} \;\,\beta_\alpha = \beta \;\,{\rm and}\;\,\mu_\alpha = \mu.
\end{equation}
Here, $\ZZ\!\left (\beta,\H,\mu,\hat{N}\right ) = \Tr[\ee^{-\beta\left (\H-\mu \hat{N}\right )}]$ is the partition function. 

On the other hand, if the reservoirs are initially out of equilibrium with each other, the imbalance in temperature or chemical potential sets up a current of energy or particles flowing into $S\!$ from $B_\alpha$. These energy and particle currents are respectively denoted $J^E_\alpha$ and $J^P_\alpha$, while $J^Q_\alpha = J^E_\alpha - \mu_\alpha J^P_\alpha$ is the corresponding heat current. In the stationary state, the second law of thermodynamics requires that \textit{the total rate of entropy production in the reservoirs is non-negative}:
\begin{equation}
\label{secondLaw}
-\sum_\alpha \beta_\alpha J^Q_\alpha \geq 0.
\end{equation}

Assuming that different reservoirs couple to different regions of the network, basic conservation laws imply that the transfer of energy or particles between the reservoirs can only occur via commensurate energy or particle currents flowing within the system. Therefore, there exist one or more current observables on $S\!$, here denoted schematically by $\hat{J}_{S\!}$, having non-zero expectation value in the NESS:
\begin{equation}
\label{currentSystem}
\avg{\hat{J}_{S\!}}_\infty \neq 0 \quad {\rm unless} \quad J^{E,P}_\alpha = 0.
\end{equation}
(We denote expectation values at time $t$ by $\langle \bullet \rangle_t$, with $\langle \bullet \rangle_\infty$ the limiting value as $t\to \infty$.) The existence of such internal currents in a boundary-driven system implies that \textit{the non-equilibrium steady state must exhibit coherence} in the eigenbasis of $\H_{S\!}$. Although we defer a detailed demonstration and discussion of this claim to Appendix~\ref{app_coherenceCurrents}, its plausibility can be appreciated by considering the example of a one-dimensional (1D) network with open (i.e.~non-periodic) boundary conditions. For this geometry, the eigenstates of $\H_{S\!}$ do not support internal currents at all~\cite{Wichterich2007pre}. This follows because, in the absence of external sources or sinks, any such current would lead to an accumulation of particles or energy in one part of the system, which is incompatible with the fact that energy eigenstates are stationary states of the closed-system dynamics.

A widely used dynamical model of the situation depicted in Fig.~\ref{fig_openNetwork} is the quantum master equation
\begin{equation}
\label{genericME}
\partial_t\r_{S\!}(t) = -\ii \left [\H_{S\!},\r_{S\!}(t)\right ] + \LL \r_{S\!}(t),
\end{equation}
valid for times $t$ much greater than the environment memory time. In order for Eq.~\eqref{genericME} to generate a completely positive and trace-preserving (CPTP) evolution for any state $\r_{S\!}(t)$, the dissipator $\LL$ must be in Lindblad form~\cite{Lindblad1976cmp,Gorini1976jmp} $\LL = \sum_{j} \gamma_{j} \DD[\hat{L}_{j}]$, where $\DD[\hat{L}]\r_{S\!} = \hat{L}\r_{S\!}\hat{L}^\dagger - \tfrac{1}{2}  \lbrace \hat{L}^\dagger\hat{L},\r_{S\!}  \rbrace$ and $\hat{L}_{j}$ is a jump operator describing an incoherent transition occurring at a rate $\gamma_{j}$.

Since the reservoirs are assumed to initially be  statistically independent, the standard construction of the dissipator is a sum of generators $\LL_\alpha$ representing each bath $B_\alpha$, i.e.
\begin{equation}
\label{additive}
\LL = \sum_\alpha \LL_\alpha.
\end{equation}
This constitutes our definition of additivity, as studied previously in Refs.~\onlinecite{Chan2014pra,Giusteri2017pre, Brask2017arX}. This is distinct from the concept of additive decoherence rates explored, for example, in Refs.~\onlinecite{Yu2006prl,Lankinen2016pra}. The additivity assumption permits one to unambiguously identify the particle current $J^P_\alpha(t)$ and energy current $J^E_\alpha(t)$ entering the system from $B_\alpha$ as
\begin{align}
\label{currents}
J^P_\alpha(t) = \avg{\LL_\alpha^\dagger \hat{N}_{S\!}}_t, \quad J^E_\alpha(t) = \avg{\LL_\alpha^\dagger \H_{S\!}}_t.
\end{align}
Here, $\LL_\alpha^\dagger$ is the adjoint generator describing the Heisenberg-picture evolution of observables, defined by $\Tr[\hat{B}\LL_\alpha^\dagger\hat{A}] =\Tr[\hat{A}\LL_\alpha\hat{B}]$ for arbitrary operators $\hat{A}$ and $\hat{B}$.

Regarding the specific form of the generators $\LL_\alpha$, various inequivalent approaches are commonly employed. These can be broadly classified into two groups according to the fixed point of each generator, i.e.\ the state $\hat{r}_\alpha$ satisfying $\nolinebreak{\LL_\alpha \hat{r}_\alpha = 0}$. The first is the ``global'' approach, where each generator drives the entire network towards the corresponding equilibrium state, i.e.\
\begin{equation}
\label{globalFixedPoint}
\LL_\alpha\hat{r}_\alpha = 0\Longleftrightarrow\hat{r}_\alpha = \frac{\ee^{-\beta_\alpha \left (\hat{H}_{S\!}-\mu_\alpha\hat{N}_{S\!}\right )}}{\ZZ\!\left (\beta_\alpha,\H_{S\!},\mu_\alpha,\hat{N}_{S\!}\right )}.
\end{equation}
Note that here we assume the fixed point $\hat{r}_\alpha$ to be unique. 

Lindblad generators satisfying~\eqr{globalFixedPoint} can be derived from a microscopic model under the Born-Markov and secular approximations~\cite{BreuerPetruccione}. The latter approximation is assumed to be justified in the quantum-optical regime, where the separation between energy levels of $\H_{S\!}$ is much larger than their environment-induced broadening. \eqr{globalFixedPoint} directly implies the correct equilibration behaviour~\eqref{rhoInfEquilibrium}, and also ensures---via the Spohn inequality~\cite{Spohn1978jmp}---that the second law~\eqref{secondLaw} holds for the heat currents from the baths. However, generators derived under the aforementioned approximations also have the property that they do not couple populations and coherences in the eigenbasis of $\H_{S\!}$~\cite{BreuerPetruccione}. Adding such generators together according to \eqr{additive} therefore generates independent equations of motion for the populations and coherences. Since the evolution is trace-preserving, if the stationary state is unique then it must be diagonal in the eigenbasis of $\H_{S\!}$, which is generally inconsistent with the condition~\eqref{currentSystem}~\cite{Wichterich2007pre}. 

Alternatively, one can use a ``local'' approach, where each generator $\LL_\alpha$ acts non-trivially only on one part of the network $s_\alpha\subset S\!$, driving it towards thermal equilibrium while leaving its complement $\bar{s}_\alpha$ unaffected. That is,
\begin{equation}
\label{localFixedPoint}
\LL_\alpha\hat{r}_\alpha = 0\Longleftrightarrow\hat{r}_\alpha = \frac{\ee^{-\beta_\alpha \left (\hat{H}_{s_\alpha}-\mu_\alpha\hat{N}_{s_\alpha}\right )}}{\ZZ\!\left (\beta_\alpha,\hat{H}_{s_\alpha}, \mu_\alpha,\hat{N}_{s_\alpha}\right )}\hat{O}_{\bar{s}_\alpha},
\end{equation}
where $\hat{H}_{s_\alpha}$ and $\hat{N}_{s_\alpha}$ are respectively the Hamiltonian and particle number operator of $s_\alpha$, while $\hat{O}_{\bar{s}_\alpha}$ is an arbitrary density operator with support on the complement $\bar{s}_\alpha$. Clearly, $\hat{r}_\alpha$ is not unique in this case. 

Generators satisfying~\eqr{localFixedPoint} may either be derived microscopically using various approximations~\cite{Scala2007jpa,Wichterich2007pre,Rivas2010njp,Trushechkin2016epl}, or directly postulated on phenomenological grounds~\cite{Prosen2009jsm,Linden2010prl,Ajisaka2012prb,Guimaraes2016pre}. In the local approach, $\r_{S\!}^\infty$ is not necessarily diagonal in the energy eigenbasis, and therefore provides a consistent model for the internal current dynamics as required by Eq.~\eqref{currentSystem}. However, neither~\eqr{rhoInfEquilibrium} nor~\eqr{secondLaw} hold, in general, leading to potential violations of thermodynamic laws~\cite{Levy2014epl,Stockburger2017fp}.

In what follows, we consider the simplest case of a two-site network with two independent reservoirs labelled by the index $\alpha = L,R$. We show that, in the	limit of weak system-reservoir coupling, the asymptotic dynamics is governed by a master equation whose dissipator takes the form
\begin{equation}
\label{LnonAdditive}
\LL = \LL_L + \LL_R + \LL_{\rm int}.
\end{equation}
In the quantum-optical limit, $\LL_L$ and $\LL_R$ determine the currents from the baths according to Eq.~\eqref{currents} and induce thermalisation according to Eq.~\eqref{globalFixedPoint}. Outside of the quantum-optical limit, these generators drive $S\!$ towards the reduction of a global thermal state (i.e.\ including the baths and the system-bath interaction) that accounts for system-reservoir correlations~\cite{Gogolin2016rpp}. The interference term $\LL_{\rm int}$, which appears whenever $B_L$ and $B_R$ are not in equilibrium with each other, generates coherence in eigenbasis of $\H_{S\!}$ as required by condition~\eqref{currentSystem}. In this way, all three properties~\eqref{rhoInfEquilibrium}--\eqref{currentSystem} can be satisfied, but only by abandoning the additivity assumption~\eqref{additive}. 

Physically, non-additivity stems from correlations between the two reservoirs~\cite{Chan2014pra}. Out of equilibrium, such correlations grow steadily in time due to the particle current flowing between the baths. Hence, Eq.~\eqref{additive}, which is justified by the initial statistical independence of the reservoirs, fails to hold as $t\to \infty$, even if the reservoirs are spectrally unstructured and coupled arbitrarily weakly to the system.


\section{Exactly solvable model}
\label{sec_model}

\subsection{Description of the model}
\label{sec_modelDescription}

\begin{figure}[t]\centering
\flushleft(a)\vspace{2.5mm}
\begin{minipage}{\linewidth}
\includegraphics[width=\linewidth]{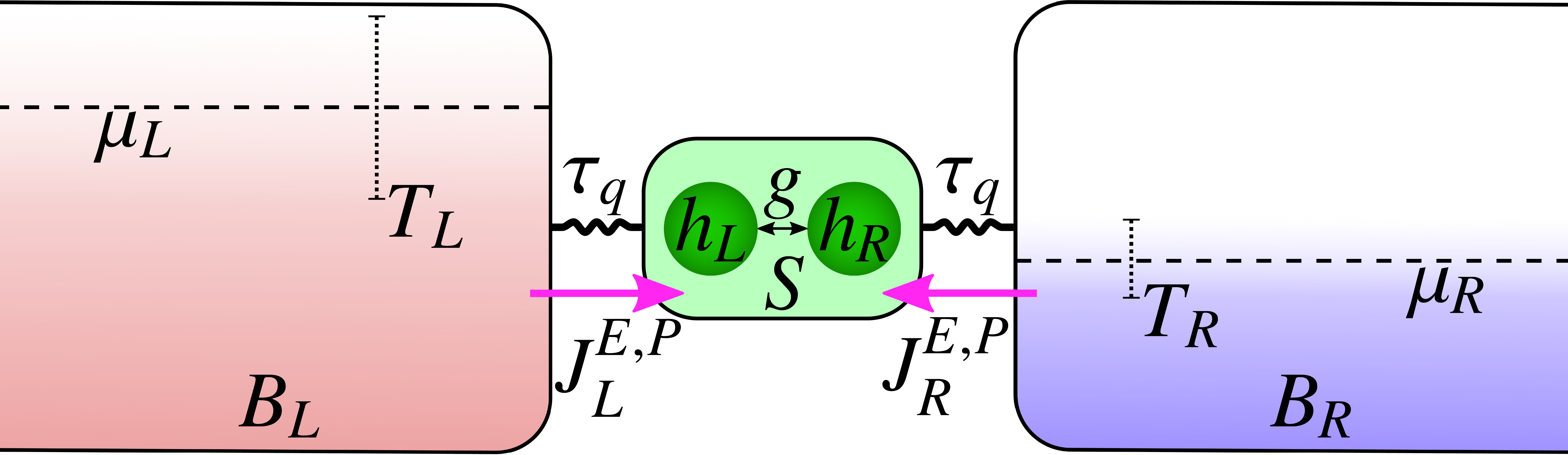}
\end{minipage}
\flushleft(b)\vspace{1mm}
\begin{minipage}{\linewidth}
\includegraphics[width=\linewidth, trim = 0mm  45cm 0mm 35cm, clip]{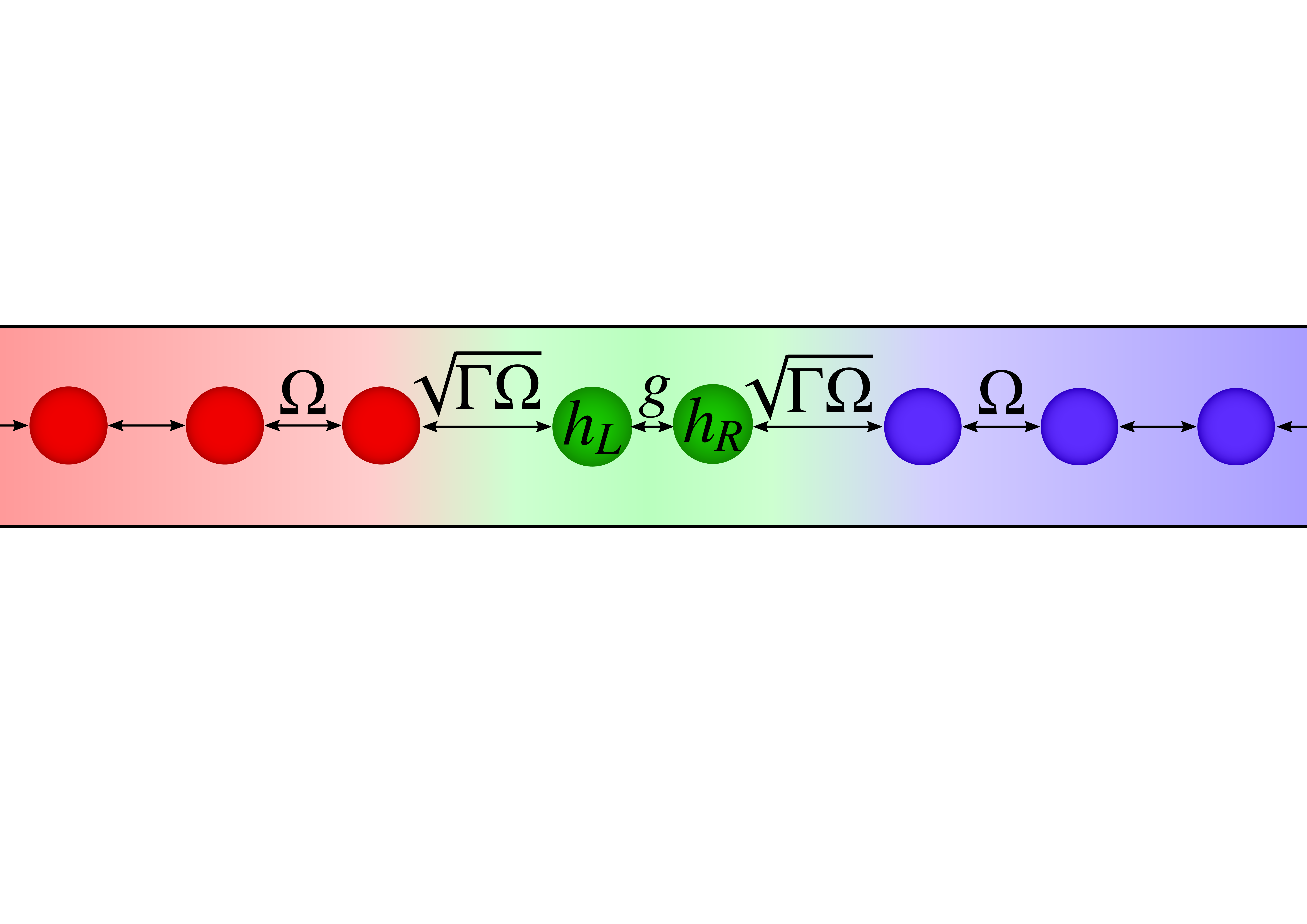}
\end{minipage}
\caption{(a)~The open system $S\!$ comprises two fermionic modes with local energies $h_{L,R}$ and tunnel coupling $g$. Each mode can exchange fermions with an independent reservoir $B_{L,R}$ at temperature $T_{L,R}$ and chemical potential $\mu_{L,R}$, thus establishing stationary energy and particle currents $J^{E,P}_{L,R}$ flowing into $S\!$. (b)~The reservoirs are explicitly modelled as 1D tight-binding chains, with $\Omega$ and$\sqrt{\Gamma\Omega}$ respectively the intra-reservoir and system-reservoir tunnelling energies.\label{fig_Model}}
\end{figure}

Let us now detail our specific set-up, which belongs to the well-known family of resonant-level transport models~\cite{Caroli1971jpc,Meir1992prl}. The total Hamiltonian takes the form $\nolinebreak{\H = \H_{S\!} + \H_B + \H_{S\!B}}$, describing a central open system $S\!$ sandwiched between two fermionic particle reservoirs $B_L$ and $B_R$, as illustrated in Fig.~\ref{fig_Model}. This models a thermoelectric tunnel junction~\cite{Benenti2017} or entangler~\cite{Brask2015njp,Tavakoli2017}, which channels a current of fermions between two conducting leads due to a temperature or chemical-potential gradient.

Specifically, $S\!$ comprises two localised fermionic modes with Hamiltonian
\begin{equation}
\label{Hs}
\H_{S\!} = \sum_{\alpha=L,R} h_\alpha \hat{n}_\alpha - \frac{g}{2}\left (\cdag_L \c_{R} + \cdag_{R} \c_L \right ),
\end{equation}
where $\c_\alpha$ annihilates a fermion on site $\alpha=L,R$ and satisfies the anti-commutation relations $\{\c_\alpha,\cdag_{\alpha'}	\} = \delta_{\alpha\alpha'}$ and $\nolinebreak{\{\c_\alpha,\c_{\alpha'}\}=0}$, while $\hat{n}_\alpha=\cdag_\alpha\c_\alpha$. We parametrise the local energies $h_\alpha$ by their mean $h=\frac{1}{2}(h_L+h_R)$ and detuning $\delta=h_L-h_R$, with $g$ the tunnel coupling between the sites. 

The baths are particle reservoirs described by the Hamiltonian $\H_B = \H_{L} + \H_{R}$. Each Hamiltonian $\H_\alpha$ describes a \textit{uniform chain}, i.e.\ a 1D tight-binding model on $M$ sites with hopping amplitude $\Omega>0$, given explicitly by
\begin{align}
\label{HrExplicit}
\H_{\alpha} & = -\frac{\Omega}{2} \sum_{m=1}^{M-1} \left ( \hat{A}_{m,\alpha}^\dagger \hat{A}_{m+1,\alpha} + \hat{A}_{m+1,\alpha}^\dagger \hat{A}_{m,\alpha} \right )\\
\label{HrFormal}
& =  \sum_{q}\omega_q \adag_{q,\alpha} \a_{q,\alpha}.
\end{align}
Here, $\hat{A}_{m,\alpha}$ annihilates a fermion localised on site $m$ of bath $B_\alpha$ and satisfies $\{\hat{A}_{m,\alpha},\hat{A}^\dagger_{m',\alpha'}\} = \delta_{mm'}\delta_{\alpha\alpha'}$ and $\{\hat{A}_{m,\alpha},\hat{A}_{m',\alpha'}\} = 0$. On the second line, the Hamiltonian is diagonalised by the canonical transformation
\begin{equation}
\label{reservoirDiag}
\a_{q,\alpha} = \sqrt{\frac{2}{M+1}}\sum_{m=1}^M \sin ( q m) \hat{A}_{m,\alpha}.
\end{equation}
The ladder operators $\a_{q,\alpha}$ describe quasi-free fermionic modes indexed by a dimensionless wave number $\nolinebreak{q=\pi k /(M+1)}$, for $\nolinebreak{k =1,2,\ldots,M}$, with dispersion relation $\nolinebreak{\omega_q = -\Omega \cos (q)}$.

The system and bath interact via tunnelling of fermions between the terminal site of each reservoir and the adjacent site of $S\!$. We parametrise the tunnelling energy as $\!\sqrt{\Gamma\Omega}$, where $\Gamma$ sets the overall frequency scale of the dissipative dynamics. Explicitly, the interaction Hamiltonian reads as
\begin{align}
\label{HsrExplicit}
\H_{S\!B} & = \frac{\sqrt{\Gamma\Omega}}{2} \sum_{\alpha=L,R} \left ( \cdag_\alpha \hat{A}_{1,\alpha} + \hat{A}_{1,\alpha}^\dagger \c_\alpha \right )\\
\label{HsrFormal}
& = \sum_{\alpha=L,R} \sum_q \left ( \tau_q \cdag_\alpha \a_{q,\alpha} + \tau_q^* \adag_{q,\alpha} \c_\alpha\right ),
\end{align}
with tunnel couplings $\tau_q = \sqrt{\Gamma\Omega/(2M+2)}\sin (q)$. 

It is convenient to analyse the problem in a basis that diagonalises $\H_{S\!}$. To do this, we collect the ladder operators into column vectors ${\bf \c} = (\c_L,\c_R)^{\mathsf{T}}$ and ${\bf \a}_q = (\a_{q,L},\a_{q,R})^{\mathsf{T}}$. We then define a new canonical set of ladder operators $\nolinebreak{{\bf \d} = (\d_1,\d_2)^{\mathsf{T}} = \mathsf{R}{\bf \c}}$ and ${\bf \b}_q = (\b_{q,1},\b_{q,2})^{\mathsf{T}}=\mathsf{R}{\bf \a}_q$, related by the orthogonal rotation matrix
\begin{equation}
\label{orth}
{\sf R} = \frac{1}{\sqrt{2\Delta}}\left ( \begin{matrix} 
\sqrt{\Delta+\delta} & -\sqrt{\Delta - \delta} \\
\sqrt{\Delta- \delta} & \sqrt{\Delta+ \delta}
\end{matrix}\right ),
\end{equation}
where $\Delta = \sqrt{g^2 +\delta^2}$. The Hamiltonian hence splits into two independent pieces $\H = \H_1 + \H_2$, with
\begin{equation}
\label{Hj}
\H_j = E_j \ddag_j\d_j + \sum_q \left (  \omega_q \bdag_{q,j} \b_{q,j} +\tau_q \ddag_j \b_{q,j} + \tau_q^* \bdag_{q,j} \d_j \right ),
\end{equation}
where $E_1 = h + \frac{1}{2}\Delta$ and $E_2 = h -\frac{1}{2}\Delta$ are the single-particle energy eigenvalues of $\H_{S\!}$. For concreteness, we assume that $E_j > 0$, or equivalently that $h_\alpha > 0$ and $\nolinebreak{ g < 2\sqrt{h_Lh_R}}$. 

The effect of the reservoirs on the system is determined by the spectral density
\begin{equation}
\label{JwDef}
\mathcal{J}(\omega) = \sum_q \lvert \tau_q \rvert^2 \delta(\omega-\omega_q).
\end{equation}
This is a smooth function of $\omega$ in the limit $M\to\infty$, where the spacing between adjacent wave vectors $\Delta q = \pi/(M+1)$ tends to zero and $q$ becomes a continuous variable taking values in the first Brillouin zone $q \in [0,\pi]$. Using the prescription $\sum_q\Delta q\to\int\dd q$, we obtain
\begin{equation}
\label{Jnewns}
\mathcal{J}_{\rm N}(\omega) = \frac{\Gamma}{2\pi} \sqrt{1 -\frac{\omega^2}{\Omega^2}}\;\Theta(\Omega-\vert\omega\vert),
\end{equation}
where $\Theta(x)$ is the Heaviside unit step function. We label this spectral density with a subscript $\rm N$ after Newns, who (to our knowledge) introduced it~\cite{Newns1969pr}. According to Eq.~\eqref{Jnewns}, each environment is characterised by a spectral bandwidth $\Omega$, leading to a vacuum correlation time of order $\Omega^{-1}$. This is rather intuitive, since $\Omega$ sets the rate at which an excitation created at the boundary of $B_\alpha$ propagates irreversibly along the chain and away from the central system's domain of influence. 

Note that choosing a 1D geometry for each environment is not as restrictive an assumption as it may appear. This is because an environment with arbitrary geometry can be mapped onto a 1D tight-binding model coupled to the system at a single boundary site, so long as the reservoir and interaction Hamiltonians take the generic forms~\eqref{HrFormal} and \eqref{HsrFormal}~\cite{Chin2010jmp,Prior2010prl,Woods2014jmp}. In general, the resulting 1D chain is described by inhomogeneous inter-site couplings and local site energies. This leads to scattering of excitations back towards the system, potentially giving rise to recurrences or other non-Markovian effects. In contrast, such backscattering is absent in the uniform chain considered here, which is characterised completely by just two frequencies $\Gamma$ and $\Omega$. 

Directly setting $\Omega^{-1}=0$ corresponds to the wide-band-limit approximation~\cite{StefanucciVanLeeuwen}, which leads to a frequency-independent spectral density. In the following, we compute the solutions for finite $\Omega$, which enables us to retain energy-dependent damping rates even for weak coupling, $\Gamma\ll \Omega$, since we need not assume that $E_j\ll\Omega$.

\subsection{Formal solution}
\label{sec_exact}

In this section, we provide the exact solution for the open-system density matrix $\r_{S\!}(t) = \Tr_B[\r(t)]$, where $\r(t)$ is the global quantum state at time $t$. The same solution formally applies to the general scenario depicted in Fig.~\ref{fig_Model}\,(a), i.e.\ any pair of environments described by Hamiltonians of the form~\eqref{HrFormal} and \eqref{HsrFormal}, corresponding to a spectral density~\eqref{JwDef}. We first describe the formal solution for this general case, before specialising to the Newns spectral density~\eqref{Jnewns} of a uniform chain in later sections. Details of the calculation are presented in Appendix~\ref{app_solutionDerivation}. 

We consider factorised initial conditions of the form $\nolinebreak{\r(0) = \r_{S\!}(0)\r_{L}\r_{R}}$, with reservoir $B_\alpha$ initialised in the Gibbs state
\begin{equation}
\label{GibbsState}
\r_{\alpha} = \frac{\ee^{-\beta_\alpha \left (\H_{\alpha} - \mu_\alpha \hat{N}_{\alpha}\right )}}{\ZZ\!\left (\beta_\alpha,\H_\alpha,\mu_\alpha,\hat{N}_\alpha\right )}.
\end{equation}
Here, $\hat{N}_\alpha = \sum_q \adag_{q,\alpha} \a_{q,\alpha}$ is the number of particles in $B_\alpha$. 

The dynamics preserves the total number of fermions due to the relation $[\H,\hat{N}]= 0$, where $\nolinebreak{\hat{N} = \hat{N}_L + \hat{N}_R + \hat{N}_{S\!}}$ and $\nolinebreak{\hat{N}_{S\!}=\hat{n}_L+\hat{n}_R}$. For any physical initial state satisfying $[\hat{N},\r(0)]= 0$, we thus have $[\hat{N}_{S\!},\r_{S\!}(t)]= \Tr_B[\hat{N},\r(t)] = 0$. Therefore, $\r_{S\!}(t)$ is characterised by five independent real numbers, in general. Four of these are encapsulated by the Hermitian correlation matrix 
\begin{equation}
\label{correlationMatrix}
C_{jk}(t) = \Tr \left [ \r_{S\!}(t) \ddag_j\d_k \right ],
\end{equation}
which satisfies $0\leq \mathsf{C}(t) \leq \mathbbm{1}$. (Here, and throughout this document, the elements of a matrix $\sf A$ are denoted by $A_{jk}$.) The fifth degree of freedom is the double-occupancy probability
\begin{equation}
\label{DofT}
D(t) = \Tr \left [\r_{S\!}(t) \ddag_1 \ddag_2 \d_2 \d_1  \right ].
\end{equation}

We compute Eqs.~\eqref{correlationMatrix} and~\eqref{DofT} by solving the equations of motion for $\d_j(t)$ and $\b_{q,j}(t)$ in the Laplace domain. Here, $\hat{O}(t) = \ee^{\ii \H t}\hat{O}\ee^{-\ii\H t}$ denotes the Heisenberg-picture time dependence of an operator $\hat{O}$. The solution for $t>0$ can be completely expressed in terms of the propagator 
\begin{align}
\label{propagatorDef}
G_{jk}(t) = \bra{0}\d_j(t)\ddag_k\ket{0},
\end{align}
where $\ket{0}$ is the vacuum state, i.e.\ $\hat{N}\ket{0} = 0$. According to Eq.~\eqref{Hj}, the particle number in each $j$ sector is separately conserved: $[\hat{H},\hat{N}_j] = 0$, with $\hat{N}_j = \ddag_j\d_j + \sum_q \bdag_{q,j}\b_{q,j}$. It thus follows from the definition~\eqref{propagatorDef} that $G_{jk}(t) = \delta_{jk}G_j(t)$, with
\begin{equation}
\label{GjExact}
G_j(t) = \int\dd\omega\,\ee^{-\ii\omega t} \varphi_j(\omega).
\end{equation}
The function $\varphi_j(\omega)$ is the probability distribution of excitation energies associated with the state $\ddag_j\ket{0}$, i.e.\
\begin{equation}
\label{phiSpectralFunction}
\varphi_j(\omega) = \sum_{n}  \lvert\bra{n}\ddag_j\ket{0} \rvert^2 \delta(\omega - \varepsilon_n),
\end{equation}
where $\ket{n}$ is a single-particle eigenstate of $\H$ with energy $\varepsilon_n$, i.e.\ $\H\ket{n} = \varepsilon_n\ket{n}$ (note that $\varepsilon_n$ may be negative~\footnote{Readers concerned by the occurrence of negative frequencies should feel reassured that this feature poses no fundamental problem because, unlike in the bosonic case, the Fermi-Dirac distribution functions $f_\alpha(\omega)$ are well behaved for $\omega < 0$. If desired, one may shift $\H \to \H + \Omega\hat{N}$ and $\mu_\alpha\to \mu_\alpha + \Omega$ without affecting the dynamics (since $[\r(t),\hat{N}]=0$), leading to slightly less wieldy equations written in terms of positive frequencies only}). Therefore,
\begin{align}
\label{normalisation}
\int\dd\omega\,\varphi_j(\omega) & = 1,\\
\label{deltaLimit}
\lim_{\Gamma\to 0} \varphi_j(\omega) & = \delta(\omega - E_j).
\end{align}
(For general spectral densities, the limit $\Gamma\to 0$ here means that all tunnel couplings $\tau_q\to 0$.)

The solution can be compactly represented in terms of the matrix $\mathsf{F}(\omega) = \mathsf{R f}(\omega)\mathsf{R^T}$, where $\mathsf{f}(\omega) = {\rm diag}[f_L(\omega), f_R(\omega)]$ with $f_\alpha(\omega) = (\ee^{\beta_\alpha(\omega-\mu_\alpha)}+ 1 )^{-1}$ the Fermi-Dirac function of reservoir $B_\alpha$, and the noise kernel
\begin{equation}
\label{noiseKernelPhi}
\mathsf{\Phi}(t) = \int\dd\omega\, \ee^{-\ii\omega t}\mathcal{J}(\omega)\mathsf{F}(\omega).
\end{equation}
With this notation, the correlation matrix reads as
\begin{align}
\label{CofTsolution}
\mathsf{C}(t) & = \mathsf{G}^{\dagger}(t) \mathsf{C}(0) \mathsf{G}(t) + \mathsf{Z}(t),\\
\label{ZofTsolution}
\mathsf{Z}(t) & = \int_0^t\dd t'\int_0^t\dd t''\,\mathsf{G}^\dagger(t')\mathsf{\Phi}(t'-t'')\mathsf{G}(t'').
\end{align}
The double-occupancy probability is
\begin{align}
\label{Dsolution}
D(t) = & \;\left \lvert\det \mathsf{G}(t)\right \rvert^{2} D(0) + \det \mathsf{Z}(t) \\ & + \,\sum_{j\neq k} \left [ \lvert G_j(t)\rvert^2  Z_{kk}(t)C_{jj}(0) - G^*_j(t)G_k(t)Z_{kj}(t)C_{jk}(0) \right]\!.\notag
\end{align}	

We see that, so long as $\lim_{t\to\infty}G_j(t) = 0$, the system relaxes to a unique stationary state $\r^\infty_{S\!}$ that is independent of the initial conditions. Moreover, this $\r^\infty_{S\!}$ is Gaussian~\cite{Dhar2012pre} because $D(\infty) = \det \mathsf{C}(\infty)$. The stationary state is therefore determined completely by its correlation matrix
\begin{align}
\label{CinfExact}
C_{jk}(\infty) =  \int\dd\omega\,\mathcal{J}(\omega)F_{jk}(\omega) &\left [\vartheta_j(\omega) + \ii\pi \varphi_j(\omega)\right ]\notag\\ \times & \left [\vartheta_k(\omega) - \ii\pi \varphi_k(\omega)\right ].
\end{align}
Here, we defined the Hilbert transform of $\varphi_j(\omega)$,
\begin{equation}
\label{varthetaDef}
\vartheta_j(\omega)  = {\rm P}\int\dd \omega'\, \frac{\varphi_j(\omega')}{\omega-\omega'},
\end{equation}
where $\rm P$ denotes the Cauchy principal value. 

The foregoing equations, which hold for any spectral density~\eqref{JwDef}, constitute the complete formal solution given knowledge of the propagator $\mathsf{G}(t)$. However, in general, computing the propagator is a challenging problem.

\subsection{Propagator for uniform-chain environments}
\label{sec_NewnsProp}

In order to find an explicit expression for the propagator, we now specialise to environments modelled by uniform chains with spectral density~\eqref{Jnewns}. Relaxation to a unique steady state is guaranteed in this case if we make the additional, technical assumption that
\begin{equation}
\label{nopoleCondition}
\sqrt{E_j^2 + \Gamma\Omega}<\Omega.
\end{equation} 
Physically, this inequality requires that the energy levels of the open system lie well within the reservoirs' energy bands, so that any initial excitation can eventually be absorbed. Under this condition, we show in Appendix~\ref{app_solutionDerivation} that
\begin{align}
\label{varphi}
\varphi_j(\omega) & = \frac{1}{1-\Gamma/\Omega}\frac{\mathcal{J}_{\rm N}(\omega)}{ (\omega - E'_j )^2 + \Gamma_j^2/4},
\end{align}
where we defined shifted energies and decay rates
\begin{equation}
\label{EGammaj}
E'_j = \frac{1-\Gamma/2\Omega}{1-\Gamma/\Omega} E_j, \qquad \Gamma_j = \frac{\sqrt{1-\Gamma/\Omega - E_j^2/\Omega^2}}{1-\Gamma/\Omega} \Gamma.
\end{equation}
The propagator is then evaluated using Eq.~\eqref{GjExact}. We also prove in Appendix~\ref{app_solutionDerivation} that
\begin{equation}
\label{varThetaValue}
\vartheta_j(\omega) = \left [\left (1-\frac{\Gamma}{2\Omega}\right )\omega - E_j\right ]\frac{\varphi_j(\omega)}{\mathcal{J}_{\rm N}(\omega)}.
\end{equation}

\subsection{Steady-state observables for uniform-chain environments}
\label{sec_SS}

We now present solutions for some interesting observables in the asymptotic stationary state, using the explicit formulae quoted in the previous section. If the baths are initially in equilibrium, i.e.\ $\beta_\alpha = \beta$, $\mu_\alpha = \mu$ and $f_\alpha(\omega) = f(\omega)$, then 
\begin{equation}
\label{Ceq}
C_{jk}(\infty) = \delta_{jk}\int\dd \omega\, \varphi_j(\omega) f(\omega),
\end{equation}
which differs from the strict equilibrium value $f(E_j)$ due to the finite width of the energy distribution $\varphi_j(\omega)$. Using Eqs.~\eqref{phiSpectralFunction} and~\eqref{Ceq}, we show in Appendix~\ref{app_thermalisation} that, for equilibrium baths, $\r^\infty_{S\!}$ is the reduced state of the \textit{global} Gibbs ensemble
\begin{equation}
\label{rhoInfThermalMarginal}
\r^\infty_{S\!} = \Tr_B \left [ \frac{\ee^{-\beta \left (\H-\mu \hat{N} \right )}}{\ZZ\!\left (\beta,\H,\mu,\hat{N}\right )} \right ].
\end{equation}
We thus infer that the energy-level broadening represented by Eq.~\eqref{Ceq} arises from the system-bath correlations reflected in Eq.~\eqref{rhoInfThermalMarginal}, both being related to a finite system-reservoir coupling $\Gamma$. With the help of Eq.~\eqref{deltaLimit}, we recover thermalisation in the strict weak-coupling sense~\eqref{rhoInfEquilibrium} in the limit $\Gamma\to 0$.

Out of equilibrium, particle and energy currents flow from each bath $B_\alpha$ into $S\!$. These currents are defined respectively as $\nolinebreak{\hat{J}^P_\alpha = -\ii[\H,\hat{N}_\alpha]}$ and $\hat{J}^E_\alpha = -\ii[\H,\H_\alpha]$. Fermion conservation demands that a corresponding particle current $\hat{J}^P_{S\!}$ flows between the two sites of the system, which is defined to satisfy the continuity equations, e.g.\ $\partial_t \hat{n}_L(t) = \hat{J}^P_L(t) - \hat{J}^P_{S\!}(t)$. Explicitly, we have
\begin{align}
\label{JpBath}
\hat{J}^P_{\alpha} & = \ii \sum_q \left ( \tau_q^* \adag_{q,\alpha} \c_\alpha-  \tau_q \cdag_\alpha\a_{q,\alpha} \right ),\\
\label{JeBath}
\hat{J}^E_{\alpha} & = \ii \sum_q \omega_q \left (  \tau_q^* \adag_{q,\alpha} \c_\alpha- \tau_q \cdag_\alpha\a_{q,\alpha} \right ),\\
\label{JpSystem}
\hat{J}^P_{S\!}  & = \frac{g}{2\ii}\left (\cdag_L\c_R - \cdag_R\c_L\right ).
\end{align}
Note that the mean intra-system current $\langle \hat{J}^P_{S\!}\rangle_t = g\Im[C_{12}(t)]$ vanishes identically if $[\H_{S\!},\r_{S\!}(t)]=0$. 

It follows from the definitions that, in the stationary state, the particle currents are homogeneous throughout the system, i.e.\ $J^P_{L} = J^P_{S\!} = - J^P_R$, where $J^P_\sigma = \langle\hat{J}^P_\sigma\rangle_\infty$, for $\sigma=S\!,L,R$. Likewise, the asymptotic mean energy currents $J^E_\alpha = \langle\hat{J}^E_\alpha\rangle_\infty$ satisfy $J^E_L = - J^E_R$. 

The asymptotic currents are calculated in Appendix~\ref{app_currents}. For the Newns spectral density~\eqref{Jnewns}, we obtain the standard Landauer formulae \cite{Datta}
\begin{align}
\label{JpInfExact}
J^P_L &= \int\dd\omega\, \mathcal{T}(\omega) \left [f_L(\omega) - f_R(\omega)\right ],\\
\label{JeInfExact}
J^E_L &= \int\dd\omega\, \omega \mathcal{T}(\omega) \left [f_L(\omega) - f_R(\omega)\right ],
\end{align}
with the transmission function
\begin{equation}
\label{transmissionFunction}
\mathcal{T}(\omega) = \frac{\pi g^2}{2}\varphi_1(\omega) \varphi_2(\omega).
\end{equation}
The transmission probability is proportional to the overlap between the energy distributions of the two eigenmodes of $\H_{S\!}$ [see Eq.~\eqref{phiSpectralFunction}]. Note that the Landauer formulae imply the second law~\eqref{secondLaw} for any transmission function $\mathcal{T}(\omega)\geq 0$~\cite{Benenti2017}.

\begin{figure}\centering
\includegraphics[width=0.8\linewidth]{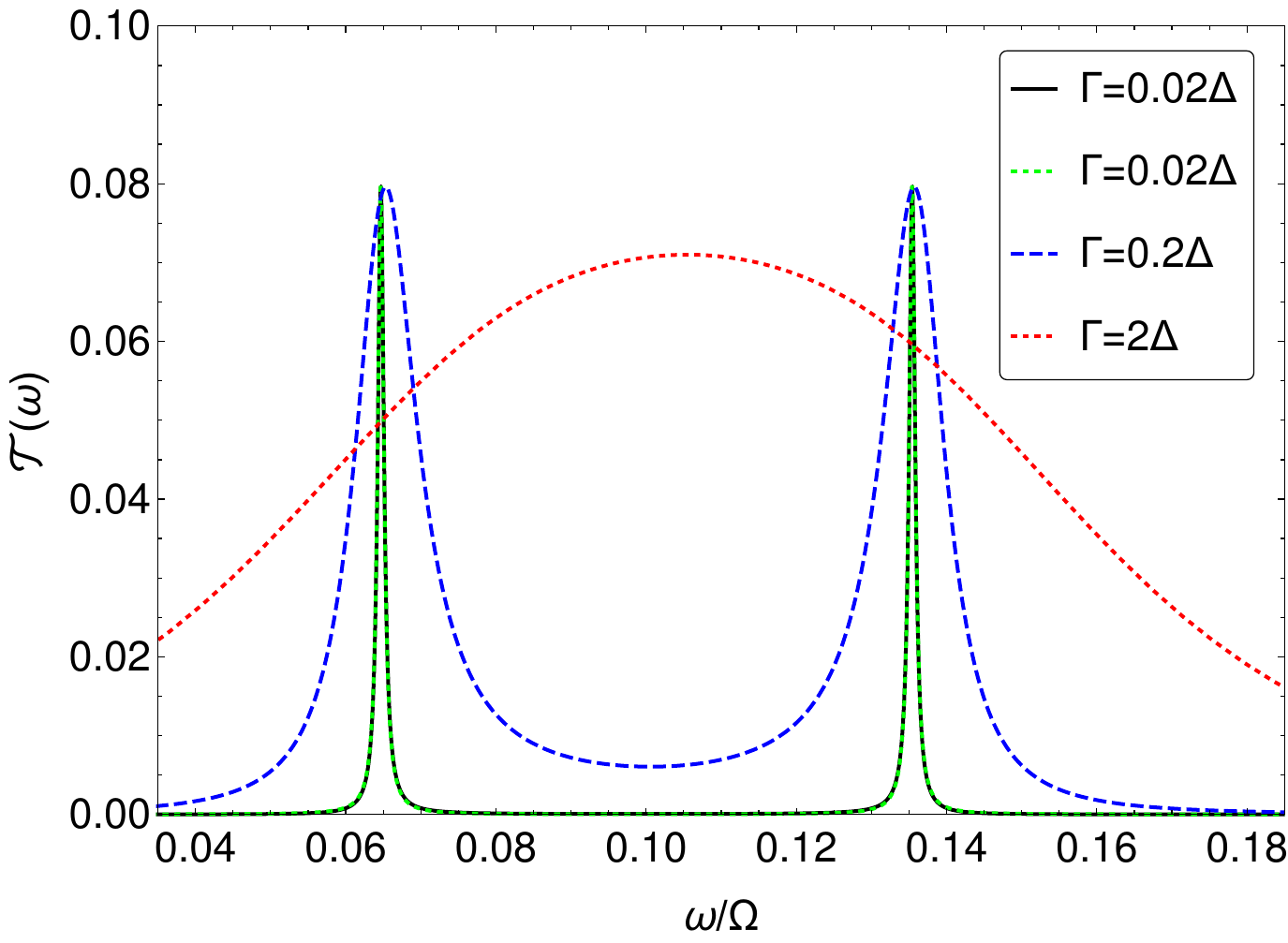}
\caption{Transmission function versus frequency, with $h=1$, $g =\delta= 0.5$ and $\Omega = 10$. The approximation \eqref{transmissionBimodal} for $\Gamma\ll\Delta$ is also shown by the green dotted line.
\label{fig_transmission}}
\end{figure}

We plot the transmission function in Fig.~\ref{fig_transmission}. For $\Gamma\ll \Delta$, $\mathcal{T}(\omega)$ is a bimodal distribution that is well approximated by\begin{equation}
\label{transmissionBimodal}
\mathcal{T}(\omega) \approx \frac{g^2}{4\Delta^2}\sum_{j=1}^2\Gamma_j\varphi_j(\omega).
\end{equation}
As $\Gamma$ is increased, the two peaks at $\omega = E_{1,2}$ broaden and ultimately merge into a single maximum for $\Gamma\gg\Delta$.


\section{Master equation}
\label{sec_ME}

\subsection{Exact master equation}
\label{sec_exactME}

In order to analyse the non-additive properties of the dynamics, we first write Eqs.~\eqref{CofTsolution} and \eqref{Dsolution} in differential form, corresponding to an exact time-local master equation for the system density operator (see Refs.~\onlinecite{Jin2010njp,Yang2013pra} for alternative derivations). All equations presented in this section hold for an arbitrary spectral density~\eqref{JwDef}. 

Assuming that $\mathsf{G}(t)$ is non-singular, we define Hermitian matrices $ \mathsf{H}(t)$ and $ \mathsf{\Gamma}(t)$ such that $\ii \mathsf{H}(t) + \tfrac{1}{2}\mathsf{\Gamma}(t) = -\mathsf{G}^{-1}	\partial_t\mathsf{G}$. We also define rate matrices $\mathsf{\Lambda}^\pm(t)$ by
\begin{align}
\label{LambdaFormalDef}
\mathsf{\Lambda}^+&= \partial_t \mathsf{Z} - \ii [\mathsf{H},\mathsf{Z}] + \tfrac{1}{2}\{\mathsf{\Gamma},\mathsf{Z}\},\notag\\
\mathsf{\Lambda}^-	 &=\mathsf{ \Gamma} - \mathsf{\Lambda}^+,
\end{align}
where time arguments are suppressed. The matrix elements of $\mathsf{\Lambda}^\pm$ correspond to the gain and loss rate coefficients appearing in the master equation, as will be seen shortly. Indeed, upon differentiating Eqs.~\eqref{CofTsolution} and \eqref{Dsolution}, some tedious algebra reveals that
\begin{align}
\label{Ceom}
\partial_t \mathsf{C} & = \ii[\mathsf{H},\mathsf{C}] -\tfrac{1}{2}\{\mathsf{\Gamma},\mathsf{C}\} + \mathsf{\Lambda}^+,\\
\label{Deom}
\partial_t D & = -\Tr[\mathsf{\Gamma}]D + \sum_{j\neq k} \left ( \Lambda^+_{jj} C_{kk} - \Lambda_{jk}^+C_{kj} \right ).
\end{align}
Direct comparison confirms that these equations of motion are equivalent to the master equation 
\begin{align}
\label{ME}
\partial_t \r_{S\!}(t) = & -\ii [\H'_{S\!}(t), \r_{S\!}(t)] + \LL(t)\r_{S\!}(t),
\end{align}
where $\H'_{S\!}(t) = {\bf\ddag} \mathsf{H}^{\mathsf{T}} \mathbf{\d}$ 
and we defined the dissipator
\begin{align}
\label{exactDissEnergy}
\LL(t) \r_{S\!} = & \sum_{j,k=1}^2 \Lambda^-_{jk}(t) \left ( \d_j\r_{S\!} \ddag_{k} - \tfrac{1}{2} \{ \ddag_{k} \d_j,\r_{S\!}  \} \right ) \notag\\ &+\, \sum_{j,k=1}^2 \Lambda^+_{jk}(t) \left ( \ddag_{k} \r_{S\!} \d_{j} - \tfrac{1}{2} \{ \d_{j} \ddag_{k},\r_{S\!}  \} \right ).
\end{align}

After diagonalising the rate matrices $\mathsf{\Lambda}^\pm(t)$ by a unitary rotation $[\mathsf{U}^{\pm}(t)]^\dagger\mathsf{\Lambda}^\pm(t)\mathsf{U}^\pm(t) = {\rm diag }[\lambda^\pm_1(t),\lambda^\pm_2(t)]$, we cast the dissipator into canonical form~\cite{Hall2014pra}
\begin{equation}
\label{dissCanonicalForm}
\LL(t) = \sum_{j=1}^2 \sum_{s=\pm} \lambda_j^s(t) \DD[\hat{L}_j^s(t)].
\end{equation}
Here, the time-dependent jump operators are defined as $\hat{L}_j^-(t) = \sum_{k}U^-_{jk}(t)\d_k$ and  $\hat{L}_j^+(t) = \sum_{k}[U^{+}_{jk}(t)]^*\ddag_k$. These describe loss and gain of excitations from and into modes determined by the eigenbases of the matrices $\mathsf{\Lambda^\pm}(t)$. Note that only if $[\mathsf{\Lambda}^+(t),\mathsf{\Lambda}^-(t)] = 0$ can we choose $\hat{L}_j^+(t) = [\hat{L}_j^-(t)]^\dagger$, i.e.\ the loss and gain modes differ, in general.

\subsection{Exponential-propagator approximation}

\label{sec_EPA}

Throughout the rest of the paper, we specialise to the Newns spectral density~\eqref{Jnewns} and assume that $\Gamma\ll \Omega$. Although the latter condition is usually deemed necessary for the Born-Markov approximation to hold, we shall see that it is by no means sufficient. 

In order to simplify the subsequent discussion, we introduce an approximation scheme where terms of order $O(\Gamma/\Omega)$ are neglected. This amounts to the replacement 
\begin{align}
\label{Gepa}
G_j(t) \approx  \ee^{-\ii E'_j t - \Gamma_j t/2}.
\end{align}
For the sake of clarity and concision, we henceforth refer to this as the exponential-propagator approximation (EPA). The EPA is justified in Appendix \ref{app_epaPropagator}, where we give an explicit expression for the error in $G_j(t)$ thus incurred. We also derive a rigorous upper bound on the magnitude of this error that is proportional to $\Gamma/\Omega$ and decays to zero as $t\to \infty$. To the same order of approximation, we write $\nolinebreak{E'_j\approx E_j}$ and $\nolinebreak{\Gamma_j = 2\pi\mathcal{J}_{\rm N}(E_j)}$. 

\begin{figure}
\begin{minipage}{0.48\linewidth}
\flushleft\scriptsize(a)
\end{minipage}
\hspace{2mm}
\begin{minipage}{0.48\linewidth}
\flushleft\scriptsize(b)
\end{minipage}\\
\begin{minipage}{0.48\linewidth}
\includegraphics[width=\linewidth]{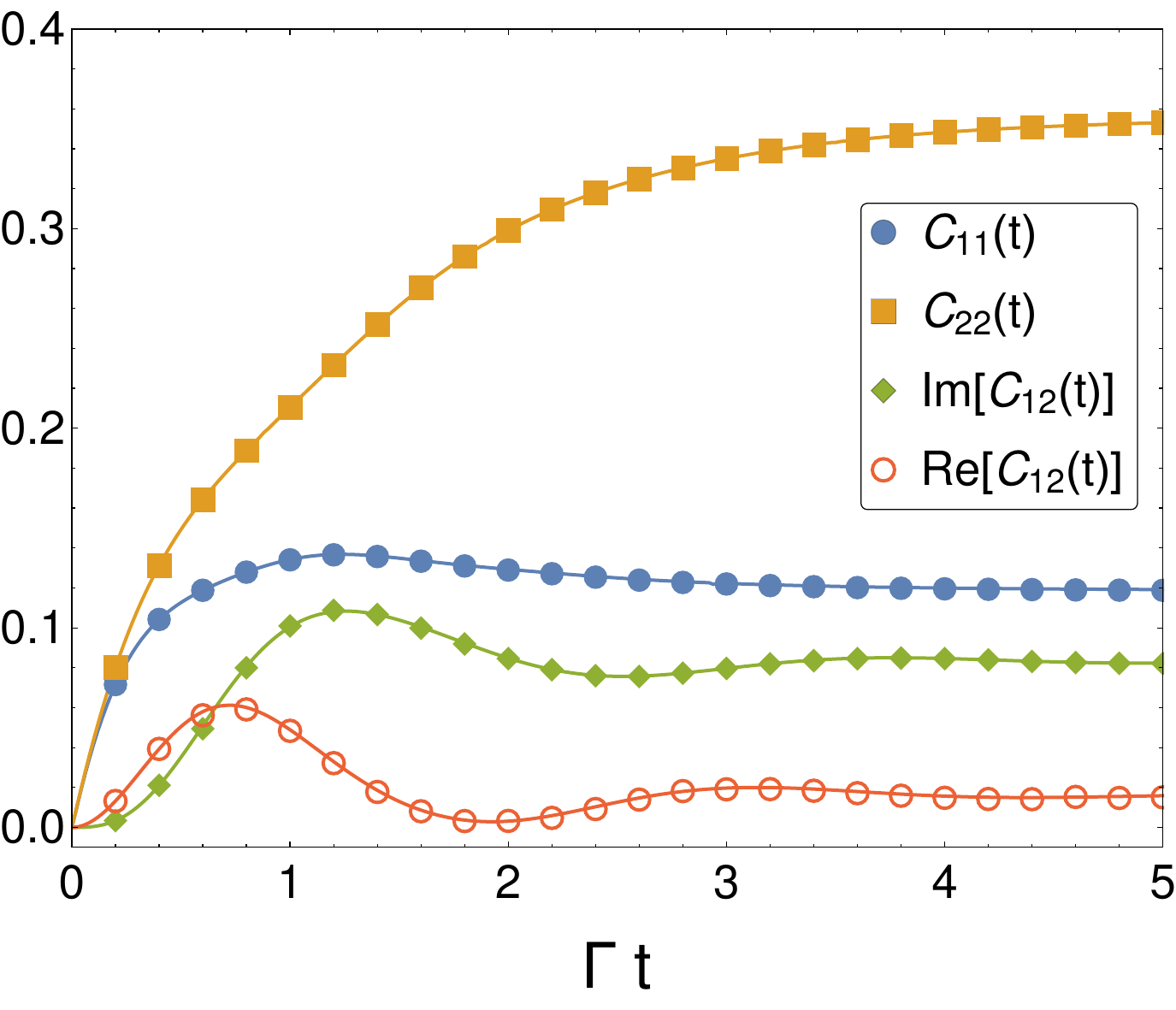}
\end{minipage}
\hspace{2mm}
\begin{minipage}{0.48\linewidth}
\includegraphics[width=\linewidth]{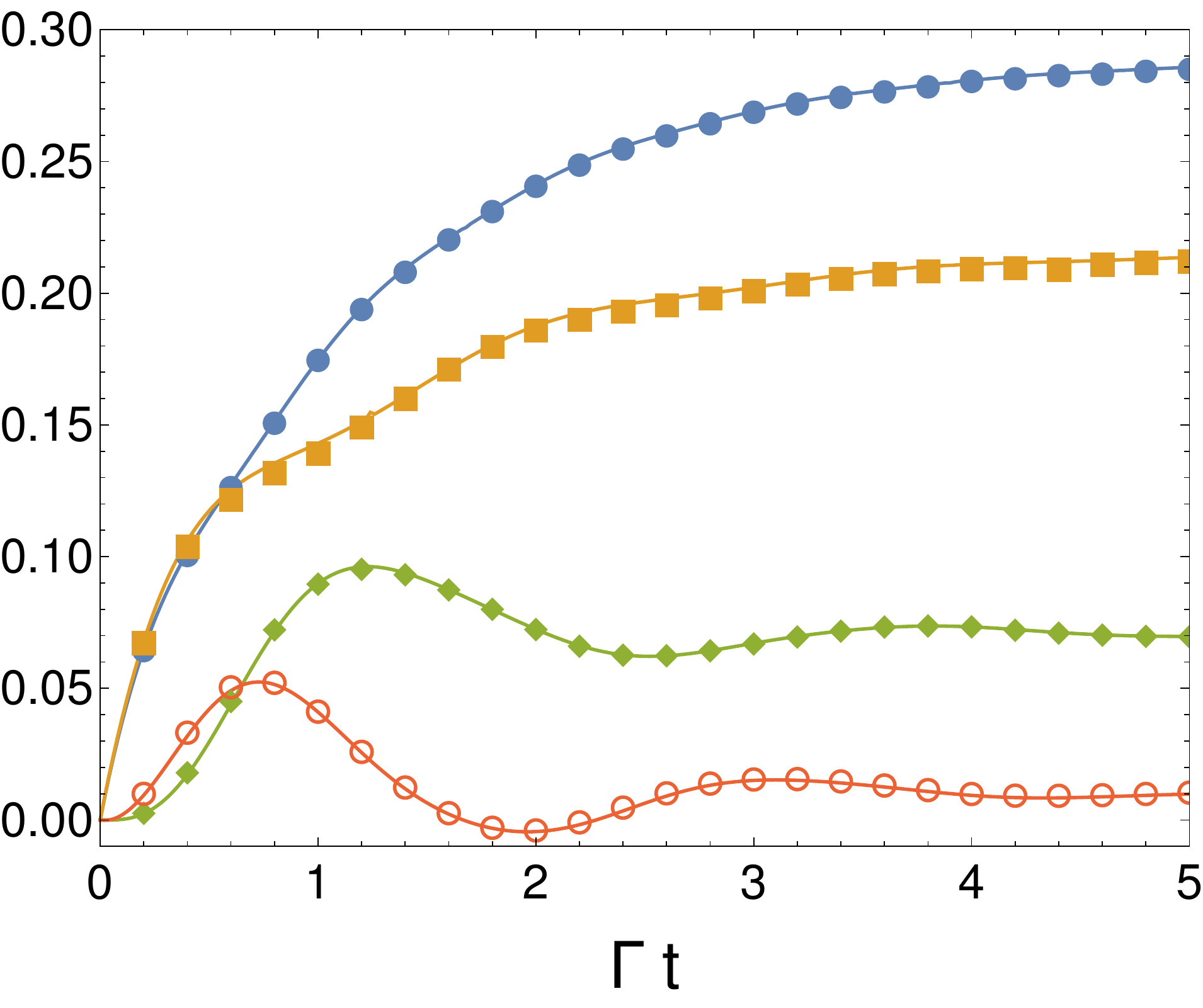}
\end{minipage}
\caption{Example evolution of the system correlation matrix with a vacuum initial condition $C_{jk}(0) = 0$, comparing exact values (points) with the EPA (lines). Parameters: $\Gamma = 0.2$, $\Omega = 100$, $\delta =0.1$, $g=0.5$; (a)~$h = 1$, $\mu_L = 1$, $\mu_R = 0$, $T_L =T_R = 0.1$; (b)~$h = 50$, $\mu_L = \mu_R = 49$, $T_L =10$, $T_R = 0.01$.
\label{fig_Ccomparison}}
\end{figure}

Unfortunately, we have not been able to derive a bound on the error induced by calculating general expectation values such as the correlation matrix~\eqref{CofTsolution} within the EPA. Nevertheless, a direct numerical comparison shows that, for sufficiently small $\Gamma/\Omega$, the EPA gives an excellent approximation to both the transient and steady-state dynamics, even if $E_j$ is comparable to $\Omega$. We illustrate the agreement between the approximation and the exact solution in Fig.~\ref{fig_Ccomparison} for a few example parameters.

We emphasise that the EPA only requires that the coupling $\Gamma$ is weak in comparison to the environment's energy scale $\Omega$, so that the approximation~\eqref{Gepa} becomes exact in the wide-band limit $\Omega\to \infty$. On the other hand, the relation between $\Gamma$ and the system energy scales $E_j$ is not restricted. 

The master equation \eqref{ME} takes a simple form under the EPA. In particular, we have that $\mathsf{H} = {\rm diag}[E_1,E_2]$, i.e.\ $\nolinebreak{\H_{S\!}'(t) = \H_{S\!}}$, $\mathsf{\Gamma} = {\rm diag}[\Gamma_1,\Gamma_2]$ and 
\begin{align}
\label{LambdaPlusEPA}
\mathsf{\Lambda}^+(t)  = \int_0^t\dd t'\, \mathsf{G}^\dagger(t')\mathsf{\Phi}(t') + \rm h.c.
\end{align}

The loss and gain modes defined by Eq.~\eqref{dissCanonicalForm} can be identified by inspection of Eq.~\eqref{LambdaPlusEPA} in two special cases. First, assume that the two baths are in thermal equilibrium with each other, so that $f_L(\omega) = f_R(\omega) = f(\omega)$. In that case, $\mathsf{F}(\omega) = f(\omega)\mathbbm{1}$ is proportional to the identity, and $\mathsf{\Lambda}^\pm(t)$ are both diagonal. Hence, $\hat{L}^-_j = (\hat{L}_j^+)^\dagger = \d_j$, i.e.\ excitations are pumped into eigenmodes of $\H_{S\!}$, driving the system towards a state that is diagonal in the energy eigenbasis of $S\!$. If the system-environment coupling is sufficiently weak, this ensures the proper thermalisation behaviour (see Section \ref{sec_asymptotic}).

Second, consider the white-noise limit, with $\Omega\to\infty$ and $\mathsf{F}(\omega) = \mathsf{F}$ a constant matrix. We then have that $\mathsf{\Phi}(t) = \Gamma\mathsf{F}\delta(t)$ and $\mathsf{\Gamma} = \Gamma\mathbbm{1}$, and since $\mathsf{G}(0) = \mathbbm{1}$ it follows that $\mathsf{\Lambda}^+(t) = \Gamma\mathsf{F}$ and $\mathsf{\Lambda}^-(t) = \Gamma(\mathbbm{1}-\mathsf{F})$. Therefore, the rate matrices $\mathsf{\Lambda}^\pm$ can be diagonalised by the rotation $\mathsf{R}^{\mathsf{T}}$, leading to Lindblad operators $\hat{L}^-_j = (\hat{L}_j^+)^\dagger$ with $\hat{L}^-_{1,2} = \c_{L,R}$, i.e.\ energy is pumped into localised modes. Of course, this only holds exactly in the unphysical scenario of infinite energy density in the baths, corresponding to $\beta_j\to 0$ or $\beta_j,\lvert\mu_j\rvert\to \infty$. Nevertheless, the local master equation may be an excellent approximation for sufficiently large chemical-potential bias or temperature.

According to Eq.~\eqref{LambdaPlusEPA}, the rate matrices are determined by the product of the propagator $\mathsf{G}(t)$---which is diagonal in the energy eigenbasis---and the noise kernel $\mathsf{\Phi}(t)$---which is diagonal in the site basis. Hence, in general, the gain and loss modes correspond neither to the energy eigenbasis $\d_j$ nor the site basis $\c_\alpha$, but instead lie somewhere in between. In particular, whenever the baths are not in equilibrium with each other, the rate matrices $\mathsf{\Lambda}^\pm$ are non-diagonal and the dissipation generates some coherence in the eigenbasis of $\H_{S\!}$. This can be seen, for example, in Fig.~\ref{fig_Ccomparison}, where the coherence can be comparable in magnitude to the populations and has a large imaginary part, reflecting the current flowing through the system.


\section{Asymptotic non-additivity}
\label{sec_nonadditive}

\subsection{Non-additivity in the energy eigenbasis}
\label{sec_asymptotic}

Now we demonstrate that the asymptotic dynamics as $t\to \infty$ is not additive. We write the rate matrices in the limit simply as $\mathsf{\Lambda}^\pm = \lim_{t\to\infty} \mathsf{\Lambda}^\pm(t)$. Explicitly, these have the components
\begin{align}
\label{LambdaPlusInf}
\Lambda_{jk}^+ = \ii\int\dd\omega\;&\mathcal{J}_{\rm N}(\omega) F_{jk}(\omega) \notag \\ &\times \left [\frac{1}{E_j-\omega+\ii\Gamma_j/2} - \frac{1}{E_k-\omega-\ii\Gamma_k/2}\right ],
\end{align}
and $\Lambda_{jk}^-= \Gamma_j\delta_{jk} - \Lambda_{jk}^+$. It follows that the asymptotic dissipator $\LL = \lim_{t\to\infty} \LL(t)$ admits the decomposition
\begin{equation}
\label{dissLeftRightC}
\LL = \LL_L + \LL_R + \LL_{\rm int}.
\end{equation}
Here, the generators $\LL_{\alpha}$ represent the thermalising effect of each individual bath $\alpha = L,R$ on the populations, while the interference term $\LL_{\rm int}$ describes the generation of coherence due to the combined effect of the non-equilibrium baths, as described below. 

First, let us examine the generators $\LL_\alpha$, which take the form
\begin{align}
\label{Lalpha}
\LL_\alpha & = \sum_{j=1}^2\left (\gamma_{\alpha,j}^{-}\DD[\d_j] + \gamma_{\alpha,j}^+ \DD[\ddag_j]\right ).
\end{align}
The decay and gain rates are defined by
\begin{align}
\label{gammaAlphaPlusMinus}
\gamma_{\alpha,j}^- &= \Gamma_{\alpha,j} \int\dd\omega\, \varphi_j(\omega)\left [ 1- f_\alpha(\omega)\right ],\notag \\
\gamma_{\alpha,j}^+ &= \Gamma_{\alpha,j} \int\dd\omega\, \varphi_j(\omega) f_\alpha(\omega),
\end{align}
where $\Gamma_{L,j} = \Gamma_j R_{1j}^2$ and $\Gamma_{R,j} = \Gamma_j R_{2j}^2$. 

One readily verifies that the unique fixed point $\hat{r}_\alpha$ satisfying $\LL_\alpha \hat{r}_{\alpha} = 0$ is Gaussian, with the correlation matrix
\begin{equation}
\label{rAlphaPop}
\Tr\left [\hat{r}_\alpha \ddag_j \d_k \right ] = \delta_{jk} \int\dd\omega\, \varphi_j(\omega) f_\alpha(\omega).
\end{equation}
In the quantum-optical limit where $\Gamma\ll \Delta$, this describes an approximately thermal distribution, up to corrections due to level broadening as discussed in Section~\ref{sec_SS}. In fact, using the arguments given in Appendix~\ref{app_thermalisation}, one can show that $\hat{r}_\alpha$ is the reduction of a global Gibbs state in equilibrium with the corresponding bath, i.e.
\begin{equation}
\label{rAlphaThermalMarginal}
\hat{r}_\alpha =  \Tr_B \left [\frac{\ee^{-\beta_\alpha \left (\H-\mu_\alpha \hat{N}\right )}}	{\ZZ\!\left (\beta_\alpha,\H,\mu_\alpha,\hat{N}\right )}\right ].
\end{equation}
It follows immediately that Eq.~\eqref{globalFixedPoint} is recovered as $\Gamma\to 0$.

Another interesting property of the generators $\LL_\alpha$ in the quantum-optical limit is that they accurately reproduce the steady-state currents according to Eq.~\eqref{currents}. Indeed, a direct computation yields
\begin{align}
\label{JpDissExplicit}
\avg{\!\LL^\dagger_L\hat{N}_{S\!}\!}_\infty & \! = \frac{g^2}{4\Delta^2} \!\int\dd\omega\sum_{j=1}^2\Gamma_j \varphi_j(\omega)\left [f_L(\omega) - f_R(\omega)\right ],\\
\label{JeDissExplicit}
\hspace{-3mm}\avg{\!\LL^\dagger_L\H_{S\!}\!}_\infty & \!	= \frac{g^2}{4\Delta^2} \!\int\dd\omega\sum_{j=1}^2 E_j \Gamma_j\varphi_j(\omega)\left [f_L(\omega) - f_R(\omega)\right ],
\end{align}
with $\sum_\alpha\langle \LL^\dagger_\alpha\hat{N}_{S\!}\rangle_\infty = 0 = \sum_\alpha\langle\LL^\dagger_\alpha\hat{H}_{S\!}\rangle_\infty$. These expressions can be shown to be equivalent to the exact Landauer formulae~\eqref{JpInfExact} and \eqref{JeInfExact} using Eq.~\eqref{transmissionBimodal} and writing $\nolinebreak{E_j \varphi_j(\omega) \approx \omega\varphi_j(\omega)}$, which is a valid approximation in the quantum-optical regime.

Since the generators $\LL_\alpha$ are in Lindblad form, they obey the Spohn inequality~\cite{Spohn1978jmp} 
\begin{equation}
\label{Spohn}
\Tr \left \lbrace\left [\ln \hat{r}_\alpha - \ln \r_{S\!}(t) \right ]\LL_\alpha \r_{S\!}(t)\right \rbrace \geq 0.
\end{equation}
However, $\LL_{\rm int}$ does not, by itself, generate a positive evolution, and thus does not necessarily satisfy such an inequality~\cite{MullerHermes2017ahp}. We nonetheless show in Appendix~\ref{app_entropyProduction} that, in the weak-coupling limit, from the Spohn inequality one may recover the second law of thermodynamics in the form 
\begin{equation}
\label{secondLawLimit}
-\lim_{\Gamma\to 0}\frac{1}{\Gamma} \sum_\alpha \beta_\alpha J^Q_\alpha \geq 0,
\end{equation}
with the heat current defined by $\nolinebreak{J^Q_\alpha = \langle \LL^\dagger_\alpha (\H_{S\!} - \mu_\alpha \hat{N}_{S\!})\rangle_\infty}$. Here, it is necessary to first divide by $\Gamma$ before taking the limit in order to avoid recovering the trivial equality $\lim_{\Gamma\to 0}\sum_\alpha \beta_\alpha J^Q_\alpha = 0$ implied by $\lim_{\Gamma\to 0}J^Q_\alpha = 0$. Corrections to the LHS of inequality~\eqref{secondLawLimit} for finite $\Gamma$ are quoted explicitly in Appendix~\ref{app_entropyProduction}.

Now we turn to the interference contribution, defined by
\begin{align}
\label{Lc}
\LL_{\rm int}\r_{S\!} =  \sum_{j\neq k} \Lambda^+_{jk} & \left [ \left (\ddag_k\r_{S\!}\d_j - \tfrac{1}{2}\{\d_j\ddag_k,\r_{S\!}\} \right ) \right . \notag \\ &\left . -\left (\d_j\r_{S\!}\ddag_k - \tfrac{1}{2}\{\ddag_k\d_j,\r_{S\!}\} \right )\right ].
\end{align}
Here, $\Lambda^+_{12} =\left (\Lambda^+_{21}\right)^* =  \xi+\ii \eta$, with
\begin{align}
\label{xi}
\xi & = \frac{g}{2\Delta}\int\dd\omega\,\sum_{j=1}^2 \Gamma_j\varphi_j(\omega)\left [f_L(\omega)-f_R(\omega)\right ] , \\
\label{eta}
\eta & = \frac{g}{2\Delta}\int\dd\omega\,  \left [ (E_1-\omega)\varphi_1(\omega) - (E_2-\omega)\varphi_2(\omega) \right ]\notag \\ & \hspace{15mm}\times \left [f_L(\omega)-f_R(\omega)\right ].
\end{align}
Therefore, $\LL_{\rm int}$ is associated with the difference in distribution functions $f_{\alpha}(\omega)$. In particular, $\LL_{\rm int}$ is non-negligible unless $f_L(\omega)\approx f_R(\omega)$ over the entire frequency range in which $\varphi_j(\omega)$ differs appreciably from zero.

The interference term satisfies the properties 
\begin{equation}
\label{LintPopCoh}
\LL^\dagger_{\rm int} \ddag_j\d_j = 0, \qquad \LL^\dagger_{\rm int} \ddag_1\d_2 = \Lambda^+_{12}.
\end{equation}
Hence, $\LL^\dagger_{\rm int}$ generates coherence while leaving the populations unaffected. This is to be expected, since energy-eigenbasis coherence is an intrinsic property of the NESS of a quantum network, as discussed in Section~\ref{sec_prelim}. It is noteworthy that, in the Schr\"odinger picture, $\LL_{\rm int}$ couples populations and coherences. Such a contribution is therefore precluded in the standard global Lindblad approach due to the secular approximation, which enforces decoupling of populations and coherences~\cite{BreuerPetruccione}. Nevertheless, $\LL_{\rm int}$  is generally a significant contribution even in the quantum-optical limit where the secular approximation is widely believed to be valid. We note also that $\LL_{\rm int}$ can never be written in Lindblad form. Therefore, departures from Markovian evolution can be used to detect non-additivity, as discussed in Section~\ref{sec_nonMarkovian}.

\subsection{Non-additivity in the local basis}

It is also possible to investigate the violation of additivity in a local, rather than global, picture of dissipation. We transform to the site basis $\c_\alpha$ and decompose the asymptotic dissipator as
\begin{equation}
\label{localDecomposition}
\LL = \bar{\LL}_L + \bar{\LL}_R + \bar{\LL}_{LR}.
\end{equation}
The local dissipators are Lindblad generators that act only on a single site, given by
\begin{equation}
\label{LlocalLeftRight}
\bar{\LL}_\alpha = \bar{\gamma}^-_{\alpha} \DD[\c_\alpha] + \bar{\gamma}^+_{\alpha} \DD[\cdag_\alpha],
\end{equation}
where $\bar{\gamma}^\pm_\alpha = \sum_{j=1}^2\gamma^\pm_{\alpha,j}$. The remaining contribution to $\LL$ describes delocalised, incoherent processes acting on both sites together:
\begin{align}
\label{LlrLocal}
\bar{\LL}_{LR} \r_{S\!} =  \sum_{\alpha\neq \alpha'}& \left [ \bar{\Lambda}^+_{\alpha\alpha'} \left (\cdag_{\alpha'}\r_{S\!} \c_\alpha - \tfrac{1}{2}\{\c_\alpha \cdag_{\alpha'},\r_{S\!}\} \right )\right .\notag \\   & \!\!\! +  \left .\bar{\Lambda}^-_{\alpha\alpha'} \left (\c_\alpha\r_{S\!} \cdag_{\alpha'} - \tfrac{1}{2}\{\cdag_{\alpha'} \c_\alpha,\r_{S\!}\} \right )\right ],
\end{align}
where $\bar{\Lambda}^\pm_{LR} = (\bar{\Lambda}^\pm_{RL})^*$, with
\begin{align}
\label{CrossTermRatesLocal}
\Re[\bar{\Lambda}^+_{LR}] = \, & \frac{g}{4\Delta} \int\dd\omega\, \left [ \Gamma_2\varphi_2(\omega) - \Gamma_1\varphi_1(\omega)\right ]\notag \\ 
&\hspace{11mm}\times \left [f_L(\omega)+f_R(\omega)\right ], \\
\Im[\bar{\Lambda}^+_{LR}] = \,& \frac{g}{2\Delta} \int\dd\omega\, \left [ (E_1-\omega)\varphi_1(\omega) - (E_2-\omega)\varphi_2(\omega) \right ]\notag \\ 
&\hspace{11mm}\times\left [f_L(\omega)-f_R(\omega)\right ], \\
\Re[\bar{\Lambda}^-_{LR}] = \, & \frac{g}{4\Delta} \int\dd\omega\, \left [ \Gamma_2\varphi_2(\omega) - \Gamma_1\varphi_1(\omega)\right ]\notag \\ 
&\hspace{11mm}\times\left [2 - f_L(\omega)-f_R(\omega)\right ],\\
\Im[\bar{\Lambda}^-_{LR}] = \, & -\Im[\bar{\Lambda}^+_{LR}].
\end{align}
The cross-term $\bar{\LL}_{LR}$ is associated with the difference between the frequency distributions $\varphi_j(\omega)$ of the open system's energy levels. That is, $\bar{\LL}_{LR}$ reflects the extent to which the occupation numbers $f_\alpha(\omega)$ of reservoir states differ between the distinct frequency ranges sampled by the two distributions $\varphi_j(\omega)$. In particular, $\bar{\LL}_{LR}$ is negligible only in the white-noise limit, where $\Delta\ll\Omega$ and the distribution functions $f_{L,R}(\omega)$ are essentially constant over the range in which $\varphi_j(\omega)$ are non-zero.

Note that here the local generators $\bar{\LL}_{\alpha}$ can be meaningfully associated with bath $B_\alpha$ only in the sense that each one depends only on the variables of $B_\alpha$ and acts non-trivially only on site $\alpha$ of the system. Even in the weak-coupling limit, $\bar{\LL}_{L,R}$ do not obey the properties~\eqref{currents} and \eqref{localFixedPoint} expected of additive, thermal dissipators unless $\delta \approx 0$, as discussed below.

\begin{figure*}\centering
\begin{minipage}{0.3\linewidth}
\flushleft (a) \hspace{10mm} $T_L=T_R=0.02\Delta$
\vspace{1.5mm}\\\includegraphics[width=\linewidth]{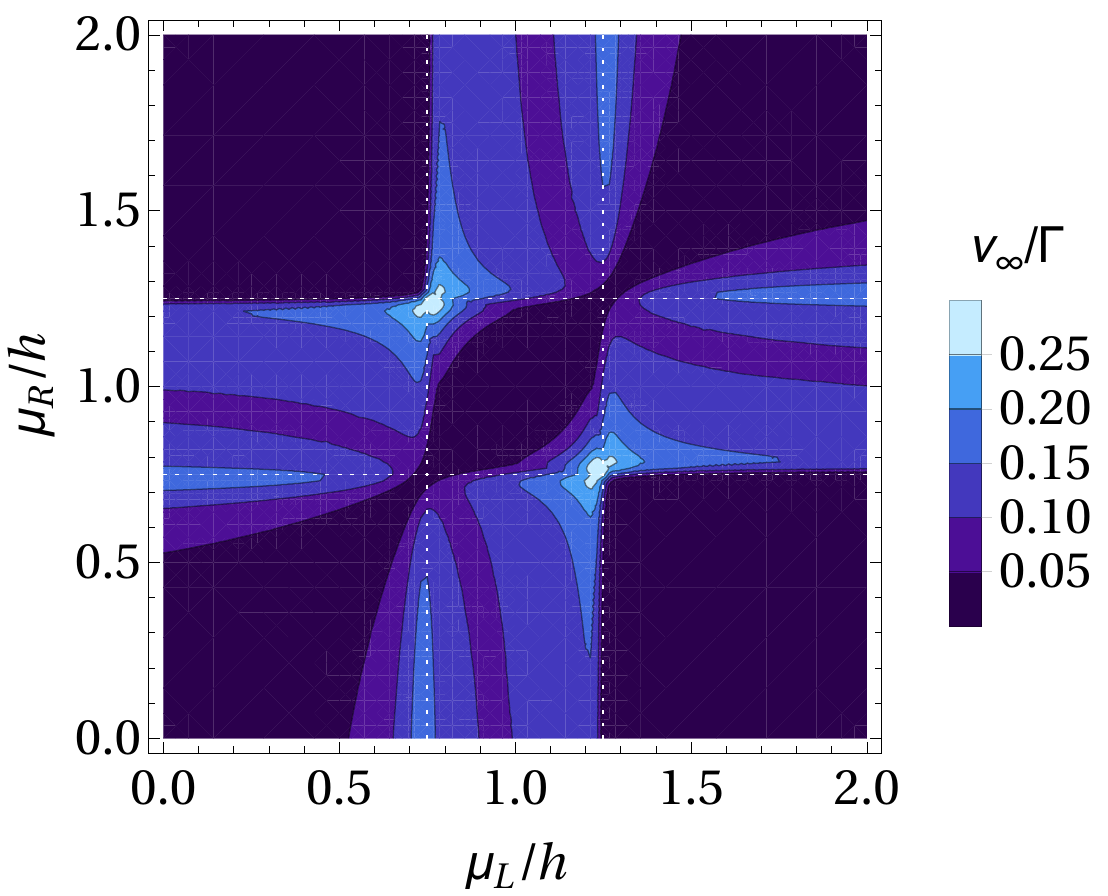}
\end{minipage}
\hspace{5mm}\begin{minipage}{0.3\linewidth}
\flushleft(b)\hspace{11mm}  $T_L=T_R = 0.2\Delta$
\vspace{1.5mm}\\\includegraphics[width=\linewidth]{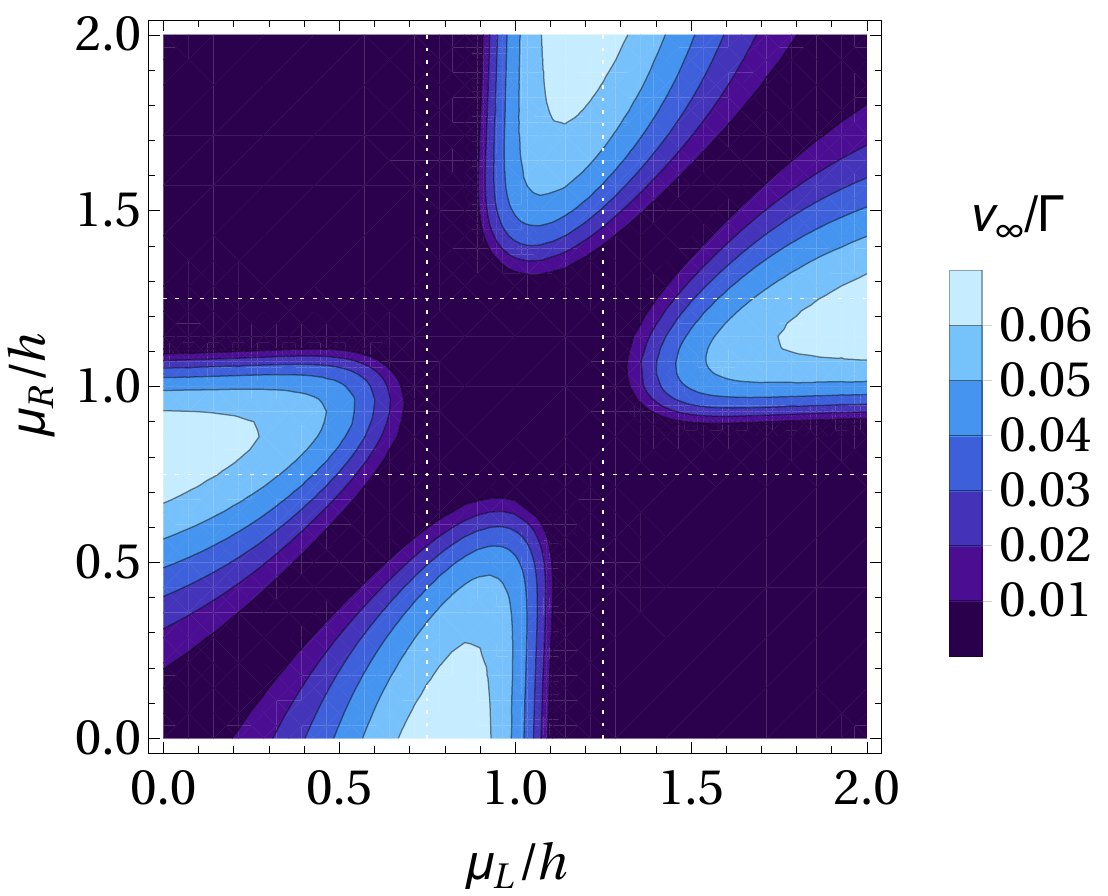}
\end{minipage}
\hspace{5mm}\begin{minipage}{0.3\linewidth}
\flushleft(c)\hspace{13mm}  $\mu_L=\mu_R = 0$
\vspace{1.5mm}\\\includegraphics[width=1.07\linewidth , trim = 0mm 0mm 0mm 5mm, clip]{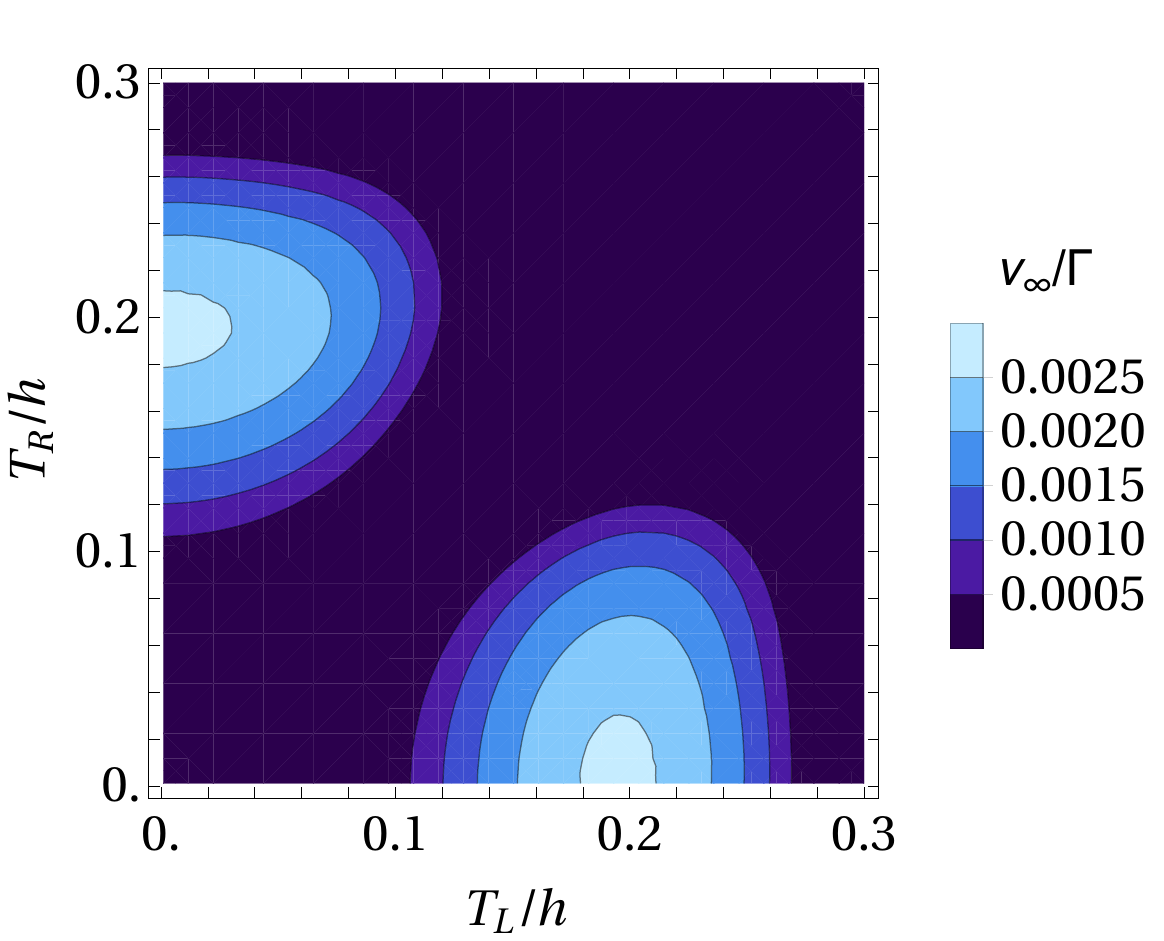}
\end{minipage}
\caption{Asymptotic non-Markovianity $u_\infty$ in the limit $\Gamma\ll\Delta\ll \Omega$, with (a), (b) variable chemical potential with fixed temperatures, or (c)~variable temperature with fixed chemical potentials. White dotted lines in (a) and (b) indicate loci where the chemical potentials $\mu_{L,R}$ equal the system energies $E_{1,2}$. Parameters: $h=1$, $\delta = 0$, $g=0.5$, $\Gamma=0.01$ and $\Omega=10$. \label{fig_NMcontoursMu}}
\end{figure*}

Let us first examine the validity of Eq.~\eqref{localFixedPoint}. The fixed point $\hat{r}_\alpha$ satisfying $\bar{\LL}_\alpha\hat{r}_\alpha = 0$ is of the form
\begin{equation}
\label{localAdditiveFixedPointEPA}
\hat{r}_\alpha = \frac{\ee^{ -\beta_\alpha \left (\H_{s_\alpha} - \mu_\alpha \hat{n}_{\alpha}\right )}}{\ZZ\!\left (\beta_\alpha,\H_{s_\alpha},\mu_\alpha,\hat{n}_{\alpha}\right )}\hat{O}_{\alpha'},
\end{equation}
where, by analogy with Eq.~\eqref{localFixedPoint}, we defined an effective local Hamiltonian acting on site $\alpha$, 
\begin{equation}
\label{effectiveLocalH}
\H_{s_\alpha} = \left [T_\alpha \ln \left (\frac{\bar{\gamma}^-_\alpha}{\bar{\gamma}^+_\alpha} \right ) +\mu_\alpha\right ]\hat{n}_\alpha,
\end{equation}
while $\hat{O}_{\alpha'}$ is an arbitrary density operator with support on the other site $\alpha'\neq \alpha$. Using Eq.~\eqref{deltaLimit}, we find 
\begin{equation}
\label{zetaBarEPA}
\lim_{\Gamma\to 0}\frac{\bar{\gamma}^-_\alpha}{\bar{\gamma}^+_\alpha}= \left [ \frac{\ee^{\beta_\alpha(h-\mu_\alpha)}+\cosh(\beta_\alpha\Delta/2)}{1+\ee^{\beta_\alpha(h-\mu_\alpha)}\cosh(\beta_\alpha\Delta/2)}\right ]\ee^{\beta_\alpha(h-\mu_\alpha)}.
\end{equation}
This only has the detailed-balance form required for thermalisation if $\beta_\alpha\Delta\ll 1$ or if $\vert\beta_\alpha(h-\mu_\alpha)\vert\gg 1$, which corresponds to the white-noise limit. In that case, we have that $\nolinebreak{\lim_{\Gamma\to 0}\ln(\bar{\gamma}^-_\alpha/\bar{\gamma}^+_\alpha) \approx \beta_\alpha(h-\mu_\alpha)}$, and the weak-coupling fixed point~\eqref{localAdditiveFixedPointEPA} can be approximated by Eq.~\eqref{localFixedPoint} with $\H_{s_\alpha} = h\hat{n}_\alpha$ and $\hat{N}_\alpha = \hat{n}_\alpha$. However, this only corresponds to the ``true'' site Hamiltonian $h_\alpha\hat{n}_\alpha$ [cf. Eq.~\eqref{Hs}] if $\delta=0$. 

Second, let us discuss the definition of the currents via Eq.~\eqref{currents}. For the particle current, we have, for example,
\begin{equation}
\label{localJpDiss}
\avg{\bar{\LL}^\dagger_L \hat{N}_{S\!}}_\infty = \avg{\LL^\dagger\hat{n}_L 	- \bar{\LL}^\dagger_{LR}\hat{n}_L}_\infty = J^P_L - \avg{\bar{\LL}^\dagger_{LR}\hat{n}_L}_\infty,
\end{equation}
where we identified $\langle\LL^\dagger \hat{n}_L\rangle_\infty = J^P_L$ using the exact master equation~\eqref{ME}. Hence, Eq.~\eqref{localJpDiss} differs from the true particle current by an amount 
\begin{equation}
\label{localJpDissCorrection}
-\avg{\bar{\LL}^\dagger_{LR}\hat{n}_L}_\infty = \frac{g}{2\Delta}\left (\Gamma_2-\Gamma_1\right )\Re \avg{\cdag_L\c_R}_\infty.
\end{equation}
Clearly, this correction is negligible if $\bar{\LL}_{LR}\approx 0$ or if $\Delta/\Omega\to 0$ so that $\Gamma_1\approx\Gamma_2$. In addition, Eq.~\eqref{localJpDissCorrection} vanishes as $\delta\to 0$, since one can easily show that $\Re \langle \cdag_L\c_R\rangle_\infty \propto \delta/\Delta$. 

On the other hand, we find for the energy current
\begin{equation}
\label{localJeDiss}
\avg{\bar{\LL}_L^\dagger\H_{S\!}}_\infty = h_L \avg{\bar{\LL}^\dagger_L \hat{N}_{S\!}}_\infty - \frac{g}{2}\sum_{j=1}^2 R_{1j}^2\Gamma_j \Re	 \avg{\cdag_L\c_R}_\infty.
\end{equation}
This expression is only correct when $\bar{\LL}_{LR} \approx 0$ and $\delta \ll h,g$, such that $\langle\bar{\LL}_L^\dagger\H_{S\!}\rangle_\infty \approx h J^P_L$. This approximately agrees with the exact Landauer formulae~\eqref{JpInfExact} and \eqref{JeInfExact} in the white-noise limit, where $\Omega\to \infty$, $\mathcal{T}(h +\omega) = \mathcal{T}(h-\omega)$ and $f_\alpha(\omega) \approx \rm const.$

\subsection{Non-Markovianity witnesses non-additivity}
\label{sec_nonMarkovian}

Since neither $\LL_{\rm int}$ nor $\bar{\LL}_{LR}$ are in Lindblad form, it may not be possible to cast the overall dissipator $\LL$ into Lindblad form for certain values of the parameters. Therefore, its canonical representation~\eqref{dissCanonicalForm} may exhibit one or more negative rates $\lambda_j^\pm(t)$ even as $t\to\infty$. Negativity of the rates $\lambda_j^s(t)$ signals the onset of non-Markovian evolution, according to the measure of non-Markovianity based on completely-positive divisible maps proposed by Rivas \textit{et al}~\cite{Rivas2010prl}. In this framework, the degree of instantaneous non-Markovianity is quantified by~\cite{Hall2014pra}
\begin{equation}
\label{nuNM}
\nu(t) = \sum_{s=\pm}\sum_{j=1,2}\max [0,-\lambda_j^s(t)].
\end{equation}

Asymptotic non-Markovianity (ANM) corresponds to $\lim_{t\to \infty}\nu(t) = \nu_\infty>0$. In a previous work, Ribeiro \textit{et al.}~\cite{Ribeiro2015prb} demonstrated that ANM arises in a class of spin and fermionic transport models, which includes the set-up considered here within the wide-band-limit approximation $\Omega^{-1}=0$. ANM has also been recently identified in several other open-system scenarios~\cite{Hall2014pra,Petrillo2016arX,Ferialdi2017pra}. 

We plot $\nu_\infty$ in Fig.~\ref{fig_NMcontoursMu}, as calculated using the EPA rate matrices defined by Eq.~\eqref{LambdaPlusInf}. At low temperatures, we find $\nu_\infty>0$ whenever the chemical potential of one reservoir lies in between the two energy eigenlevels of the system while that of the other reservoir lies outside. This effect is significantly reduced once the thermal energy grows comparable to the splitting~$\Delta$. It is remarkable that a highly non-Markovian evolution can be induced merely by tuning a macroscopic potential difference, without any engineering or microscopic control of the environment Hamiltonian. Indeed, the ANM effect survives even in the wide-band limit $\Omega\to \infty$ where each environment is completely unstructured. In Fig.~\ref{fig_NMcontoursMu}(c) we also show that some non-Markovianity may be generated by purely thermal driving, i.e.\ where the reservoirs have identical chemical potentials and different temperatures.

Since the generators $\LL_{L,R}$ ($\bar{\LL}_{L,R}$) are in Lindblad form, the appearance of asymptotic non-Markovianity is associated with the cross-term $\LL_{\rm int}$ ($\bar{\LL}_{LR}$). Hence, ANM witnesses the violation of additivity in either the global or local pictures of dissipation. Indeed, in Fig.~\ref{fig_NMcontoursMu} we observe that non-Markovianity is associated with parameters for which the master equation is not additive, in neither the local nor energy eigenbases. As discussed in previous sections, this occurs when the baths are far from equilibrium, yet the chemical-potential bias and the temperature are not so large that the white-noise limit is recovered. Since ANM is, in principle, experimentally accessible via quantum process tomography, it  represents a measurable signature of non-additivity.

\section{Conclusion}
\label{sec_conclusion}

In this work, we have presented a study of an exactly solvable microscopic system-environment model, featuring a two-site fermionic network driven out of equilibrium by two independent thermal baths. We have derived the exact master equation describing the open-system density matrix and developed a simplifying approximation scheme valid when the system-reservoir coupling $\Gamma$ is much smaller than the reservoir bandwidth $\Omega$. 

Our results demonstrate that the asymptotic master equation is not additive, i.e.\ the dissipator cannot be written as a sum of the form~\eqref{additive} where each term pertains to a single bath, even if the system-reservoir coupling is vanishingly small in comparison to every other relevant energy scale. Examining the master equation in the energy eigenbasis of the open system, we identify a generator $\LL_\alpha$ associated to each bath, which drives the open network towards thermal equilibrium according to Eq.~\eqref{globalFixedPoint} [or, more generally, Eq.~\eqref{rAlphaThermalMarginal}]. Furthermore, $\LL_\alpha$  determines the currents via Eq.~\eqref{currents} in the quantum-optical limit. This generator can therefore be interpreted as an individual contribution describing the effect of bath $B_\alpha$ acting in isolation. 

Nevertheless, additivity is violated away from equilibrium due to an additional interference term $\LL_{\rm int}$ that generates coherence, as required for a current-carrying NESS in an extended system. In extreme cases, this non-additivity can lead to an asymptotically non-Markovian master equation. Despite this, the interference term $\LL_{\rm int}$ leaves the energy-eigenbasis populations invariant and therefore the global, additive Lindblad model obtained by setting $\LL_{\rm int} = 0$ still leads to accurate approximations for the boundary currents in the quantum-optical regime. Such an additive model coincides in the limit $\Gamma\to 0$ with the standard master equation derived under the BMA and secular approximation.

We have also examined the violation of additivity in the local basis. However, here it is not always possible to identify thermal generators satisfying the properties~\eqref{currents} and~\eqref{localFixedPoint}, even in the weak-coupling limit. Thus, the identification of $\bar{\LL}_\alpha$ as an ``individual'' bath contribution is not fully justified, since it does not accurately represent the effect of a single bath acting in isolation. Nevertheless, an additive, local Lindblad model is recovered when the modes are near-resonant, i.e.\ $\delta\ll h,g$, and when the noise correlation time is much smaller than the inverse frequency splitting $\Delta^{-1}$. This conclusion is entirely consistent with recent studies showing that local additive Lindblad models may perform well in non-equilibrium scenarios when the network nodes are nearly degenerate~\cite{Gonzalez2017, Hofer2017}, but fail for large detunings~\cite{Levy2014epl}.

It has been shown~\cite{Brask2017arX} that additive open-system dynamics is obtained whenever the BMA (or, more generally, the second-order time-convolutionless projection operator method~\cite{BreuerPetruccione}) holds. Hence, our results indicate a breakdown of these assumptions at asymptotically large times when multiple thermal reservoirs act in competition. Note that, while the BMA is commonly cast in terms of an assumption that the system-reservoir density matrix factorises, it is more accurately described as a projection $\PP$ onto product states of the form
\begin{equation}
\label{TCLprojection}
\PP \r(t) = \r_{S\!}(t)\r_B,
\end{equation}
where $\r_B$ is the initial state of the environment, in combination with a low-order perturbation expansion in the system-reservoir interaction~\cite{Rivas2010njp}. 

The breakdown of additivity for arbitrarily weak system-bath coupling $\Gamma$ indicates that Eq.~\eqref{TCLprojection} does not serve as a good reference point for a perturbative expansion in powers of $\Gamma$. This can be understood as a consequence of the correlations that build up between the two parts of the environment due to the current flowing between them. As time increases, these correlations cause the joint state $\r(t)$ to move arbitrarily far away from the subspace spanned by states of the form~\eqref{TCLprojection} with $\r_B = \r_L\r_R$. A very interesting open question is whether a projection of the type~\eqref{TCLprojection} can still facilitate a good perturbative description of the long-time dynamics, but with a correlated reservoir state $\r_B$.

Although the evidence presented here pertains only to the specific system considered, we expect the qualitative conclusions regarding non-additivity to hold for other extended open systems, where currents and energy-eigenbasis coherence are inextricably linked. Indeed, previous results demonstrating asymptotic non-Markovianity in larger 1D fermionic and spin networks driven out of equilibrium~\cite{Ribeiro2015prb} already support this conclusion, because ANM witnesses non-additivity, as we have shown. Furthermore, our model can be readily generalised to a bosonic setting where the ladder operators obey commutation, rather than anti-commutation, relations. Since the Heisenberg equations are essentially identical in this case, one expects the same conclusions regarding additivity to hold, although we leave a detailed analysis of this problem to future work (see also Refs.~\onlinecite{Martinez2013prl,Freitas2017pre}). We note also a very recent study of a different bosonic model of coupled mechanical oscillators that displays coherences in the NESS~\cite{Boyanovsky2017pra}, which are naturally explained by the interference between non-equilibrium baths.

Looking ahead, it will be interesting to investigate the consequences of the results presented here for the thermodynamics of small heat machines running between multiple thermal reservoirs. For example, it remains to be seen how non-additive noise may affect the thermodynamic power and efficiency of such machines, or their ability to generate quantum resources such as coherence and entanglement. One may also ask whether the interference between the reservoirs is manifested in the fluctuations of the energy and particle currents. Finally, the framework presented herein appears to be an ideal setting to explore strong-coupling effects in thermodynamics~\cite{IlesSmith2014pra,Gallego2014njp,Esposito2015prl,Gelbwaser2015jpc,Newman2017pre,PerarnauLlobet2017}, since system-environment correlations in both equilibrium and far-from-equilibrium states may be taken into account.

\acknowledgments{We gratefully acknowledge the stimulating comments and conversation of Jonatan Brask, Nicolas Brunner, Luis Correa, John Goold, G\'eraldine Haack, Patrick Hofer, Susana Huelga, Andreas Lemmer, Fabio Mascherpa, Mart\'i Perarnau-Llobet, Ralph Silva, Andrea Smirne and Philipp Strasberg. This work was funded by the ERC Synergy grant BioQ and the EU project QUCHIP.}

\let\C\c
\let\c\cedilla
\bibliographystyle{apsrev4-1}
\bibliography{bibliography} 
\let\c\C

\appendix

\section*{Appendices}

\section{Connection between conserved currents and coherence}
\label{app_coherenceCurrents}

\begin{figure}
\includegraphics[width=0.7\linewidth, trim = 5cm 5cm 2cm 2cm, clip]{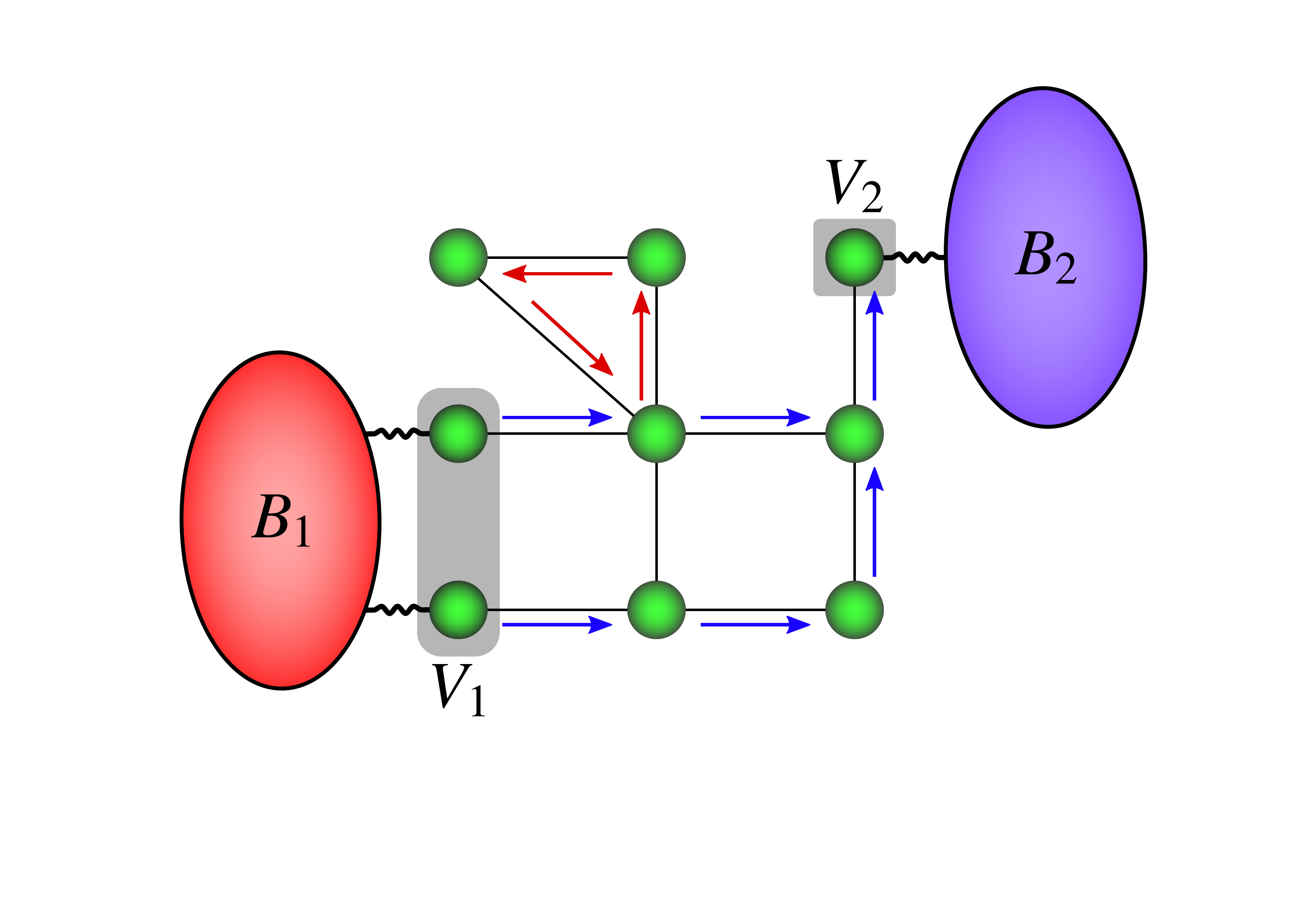}
\caption{Illustration of an open network coupled to two external reservoirs $B_{1,2}$ at the boundary regions $V_{1,2}$. Red arrows depict an internal current distribution that could satisfy the constraint~\eqref{currentConstraint}, while blue arrows represent a current distribution that must violate this constraint on the boundary lattice sites.\label{fig_networkCurrents}}
\end{figure}

In this appendix, we elucidate the relationship between conserved currents and coherence in the energy eigenbasis. Consider an open quantum network $S\!$ connected to several baths $B_\alpha$, such that the total Hamiltonian is
\begin{equation}
\label{totalHsBaths}
\H = \H_{S\!} + \sum_{\alpha} \left ( \H_{B_\alpha} + \H_{S\!B_\alpha}\right ),
\end{equation}
where $\H_{S\!}$ and $\H_{B_\alpha}$ are respectively the Hamiltonians of system and baths, while $\H_{S\!B_\alpha}$ describes the coupling between $S\!$ and $B_\alpha$. Crucially, we assume that each interaction term couples to a \textit{distinct} region on the boundary of $S\!$. We denote the corresponding interaction region by $V_\alpha$. Each $V_\alpha$ is assumed to form a proper subset of the network $S\!$. See Fig.~\ref{fig_networkCurrents} for an illustration.

Suppose that there exists a conserved quantity
\begin{equation}
\label{emDef}
\hat{X} = \hat{X}_{S\!} + \sum_\alpha\hat{X}_{B_\alpha},
\end{equation}
such that $[\H,\hat{X}] = 0$, which is also \textit{locally} conserved in the sense that $[\H_{S\!},\hat{X}_{S\!}] = [\H_{B_\alpha},\hat{X}_{B_\alpha}]=0$. It follows that 
\begin{equation}
\label{HsbMcommutatorConservation}
\partial_t \hat{X}_{B_\alpha} = \ii[\H_{S\!B_\alpha},\hat{X}_{B_\alpha}] = -\ii[\H_{S\!B_\alpha},\hat{X}_{S}].
\end{equation}
For simplicity, we assume that $\hat{X}_{S\!}$ is a one-body observable of the form
\begin{equation}
\label{MsOneBody}
\hat{X}_{S\!} = \sum_k \hat{x}_k,
\end{equation}
where the local ``charge'' $\hat{x}_k$ has support only on site $k$ of $S\!$. This scenario could describe, for example, a lattice system of conserved particles or a spin network with conserved magnetisation.

Consider now a particular site on the boundary of the system, which lies in the region $V_\alpha$ that is coupled to bath $B_\alpha$. On this site, the equation of motion for $\hat{x}_k$ takes the form of a continuity equation:
\begin{equation}
\label{boundarySiteContEq}
\partial_t \hat{x}_k = \hat{J}^X_{B_\alpha\to k} + \hat{J}^X_{S\!\to k}.
\end{equation}
The current operators defined by $\hat{J}^X_{B_\alpha\to k} = \ii [\H_{S\!B_\alpha},\hat{x}_k]$ and $\nolinebreak{\hat{J}^X_{S\!\to k} = \ii [\H_{S},\hat{x}_k]}$ describe the flow of charge  from $B_\alpha$ to site $k$ and from the other sites of $S\!$ to site $k$, respectively. In particular, the operators $\hat{J}^X_{S\!\to k}$ correspond to the internal current observables denoted schematically by $\hat{J}_{S\!}$ in Section~\ref{sec_prelim}. By definition, these observables have zero mean unless the quantum state has some coherence in the eigenbasis of $\H_{S\!}$. This follows because, by the cyclic invariance of the trace,
\begin{equation}
\label{internalCurrentExp}
\left \langle  \hat{J}^X_{S\!\to k} \right \rangle_t = \Tr \left \lbrace  \ii[\r_{S\!}(t),\H_{S\!}]\hat{x}_k \right \rbrace,
\end{equation}
which clearly vanishes if $[\H_{S\!},\r_{S\!}(t)] = 0$. However, these expectation values do not vanish whenever the NESS supports a current due to the influx of charge from the external reservoirs. Indeed, summing the expectation value of Eq.~\ref{boundarySiteContEq} over all sites in $V_\alpha$, we obtain
\begin{equation}
\label{boundaryCurrentBalance}
\left \langle \partial_t \hat{X}_{B_\alpha} \right \rangle_\infty = \sum_{k\in V_\alpha} \left \langle \hat{J}^X_{S\!\to k} \right \rangle_\infty.
\end{equation}
Here, we have used Eq.~\eqref{HsbMcommutatorConservation} and the fact that expectation values of system observables are time-independent in the NESS. Identifying $J^X_\alpha = - \langle \partial_t \hat{X}_{B_\alpha} \rangle_\infty$ as the current entering the system from $B_\alpha$, the assertion~\eqref{currentSystem} is confirmed. We conclude that boundary-driven currents imply energy-eigenbasis coherence in the NESS.

For clarity, it is useful to consider the specific case of a lattice with two-body interactions described by the generic Hamiltonian
\begin{equation}
\label{twoBodyH}
\H_{S\!} = \tfrac{1}{2}\sum_{k,l} \hat{h}_{kl},
\end{equation}
where $\hat{h}_{kl} = \hat{h}_{lk} = \hat{h}_{kl}^\dagger$ is an operator with support on sites $k$ and $l$. In this case, we have
\begin{equation}
\label{Js2bodyH}
\hat{J}_{S\!\to k}^X = \sum_{l\in A_k} \hat{J}_{l\to k}^X,
\end{equation}
where $A_k$  denotes the the neighbourhood of site $k$, i.e. all sites $l$ such that $\hat{h}_{kl}\neq 0$, while $\hat{J}_{l\to k}^X = \ii[\hat{h}_{lk},\hat{x}_k]$ denotes the current flowing from site $l$ to site $k$. According to Eq.~\eqref{internalCurrentExp}, eigenstates $\ket{E}$ of $\H_{S\!}$ obey the constraint
\begin{equation}
\label{currentConstraint}
\sum_{l\in A_k} \langle E\lvert \hat{J}_{l\to k}^X\rvert E\rangle = 0. 
\end{equation}
This constraint allows for non-zero net currents around loops in the network (see, for example, Ref.~\onlinecite{Rivas2017}), such that the local charge distribution is stationary (red arrows in Fig.~\ref{fig_networkCurrents}). However, currents induced by external sources violate the constraint~\eqref{currentConstraint} at the boundaries of the system and therefore some coherence in the energy eigenbasis is necessary to represent them (blue arrows in Fig.~\ref{fig_networkCurrents}). 

The simplest example of this principle is that of a one-dimensional chain with open (i.e.\ non-periodic) boundary conditions. Since the boundary sites of such a chain have only a single neighbour, Eq.~\eqref{currentConstraint} reduces to the identity
\begin{equation}
\label{currentConstraint1D}
\langle E\lvert \hat{J}_{k\to k+1}^X\rvert E\rangle = 0. 
\end{equation}
That is, the energy eigenstates of a chain do not support conserved currents.

In conclusion, the existence of energy-eigenbasis coherence in a boundary-driven NESS of an extended open system is a rather general consequence of conservation laws and the locality of Hamiltonian interactions. We note that one may generalise the above argument to account for more general conserved quantities, such as two-body operators of the form $\hat{X}_{S\!} = \tfrac{1}{2}\sum_{k,l} \hat{x}_{kl}$. For a network with two-body interactions, this class includes the Hamiltonian itself, which is of course always conserved. 

\section{Solution of the equations of motion}
\label{app_solutionDerivation}

Here, we provide details of the exact solution presented in Section~\ref{sec_model}. The equations of motion read as
\begin{align}
\label{dsEOM}
\ii \partial_t\d_j(t) & = E_j \d_j(t) + \sum_q \tau_q \b_{q,j}(t),\\
\label{bqEOM}
\ii\partial_t\b_{q,j}(t) & = \omega_q \b_{q,j}(t) + \tau^*_q \d_j(t).
\end{align}
These are readily solved by transforming to Laplace space~\cite{Schaller}, e.g.\ $\td_j(z) = \int_0^\infty\dd t\; \ee^{-zt} \d_j(t)$. After rearranging the resulting linear, algebraic system of equations, we transform back to the time domain to obtain 
\begin{align}
\label{dSolution}
\d_j(t) & = G_j(t)\d_j + \sum_q \tau_q K_j(\omega_q,t) \b_{q,j},\\
\label{bSolution}
\b_{q,j}(t) & = \ee^{-\ii\omega_q t}\b_{q,j} + \tau_q^* K_j(\omega_q,t)\d_j \notag \\ & \hspace{3mm} +\tau_q^*\sum_{p}\tau_p I_j(\omega_q,\omega_p,t) \b_{p,j}.
\end{align}
Here, $G_j(t)$ denotes the inverse Laplace transform of $\tilde{G}_j(z)$, where
\begin{align}
\label{GjOfZ}
\tilde{G}_j(z) &= \left [z+\ii E_j + \tilde{W}(z)\right ]^{-1},\\
\label{WofZ}
\tilde{W}(z)  & = \int\dd\omega\; \frac{\SJ(\omega)}{z+\ii\omega}.
\end{align}
We also introduced the functions
\begin{align}
\label{KjDef}
K_j(\omega,t) &= -\ii\int_0^t\dd t'\,\ee^{-\ii\omega (t-t')}G_j(t'),\\
I_j(\omega,\omega',t) & = -\ii\int_0^t\dd t'\,\ee^{-\ii\omega (t-t')}K_j(\omega',t').
\end{align}
By inspection of Eq.~\eqref{dSolution} one sees that $\mathsf{G}(t)$ is indeed given by Eq.~\eqref{propagatorDef}. Expressions~\eqref{CofTsolution} and~\eqref{Dsolution}  for the correlation matrix $\mathsf{C}(t)$ and double occupancy $D(t)$ follow directly upon plugging Eq.~\eqref{dSolution} into the definitions~\eqref{correlationMatrix} and \eqref{DofT}. Note that the above equations hold for an arbitrary spectral density $\mathcal{J}(\omega)$, which determines the propagator via Eqs.~\eqref{GjOfZ} and~\eqref{WofZ}.

Let us now compute the propagator for the specific problem at hand, with spectral density~\eqref{Jnewns}. The integral~\eqref{WofZ} evaluates to	
\begin{equation}
\label{WofZexplicit}
\tilde{W}(z) = \frac{\Gamma}{2\Omega}\left (\sqrt{z^2+\Omega^2} - z \right ),
\end{equation}
where the positive (negative) sign for the square root is used for $\Re z > 0$ ($\Re z < 0$). This implies that any poles of $\tilde{G}_j(z)$ lie on the imaginary axis, which would correspond to undamped oscillations in the time domain. In order to avoid this behaviour, we assume that inequality~\eqref{nopoleCondition} holds so that $\tilde{G}_j(z)$ has no poles. As a result, $\tilde{G}_j(z)$ is holomorphic everywhere in the complex $z$ plane except for along a cut connecting the branch points at $z=\pm \ii \Omega$. Therefore, $G_j(t)\to 0$ as $t\to \infty$ and $S\!$ relaxes to a unique stationary state (see Appendix~\ref{app_epaPropagator}).

\begin{figure}
\begin{minipage}{0.48\linewidth}
\flushleft(a)
\end{minipage}
\hspace{2mm}
\begin{minipage}{0.48\linewidth}
\flushleft(b)
\end{minipage}\\
\begin{minipage}{0.48\linewidth}
\includegraphics[width=\linewidth, trim =35cm 20cm 35cm 20cm, clip]{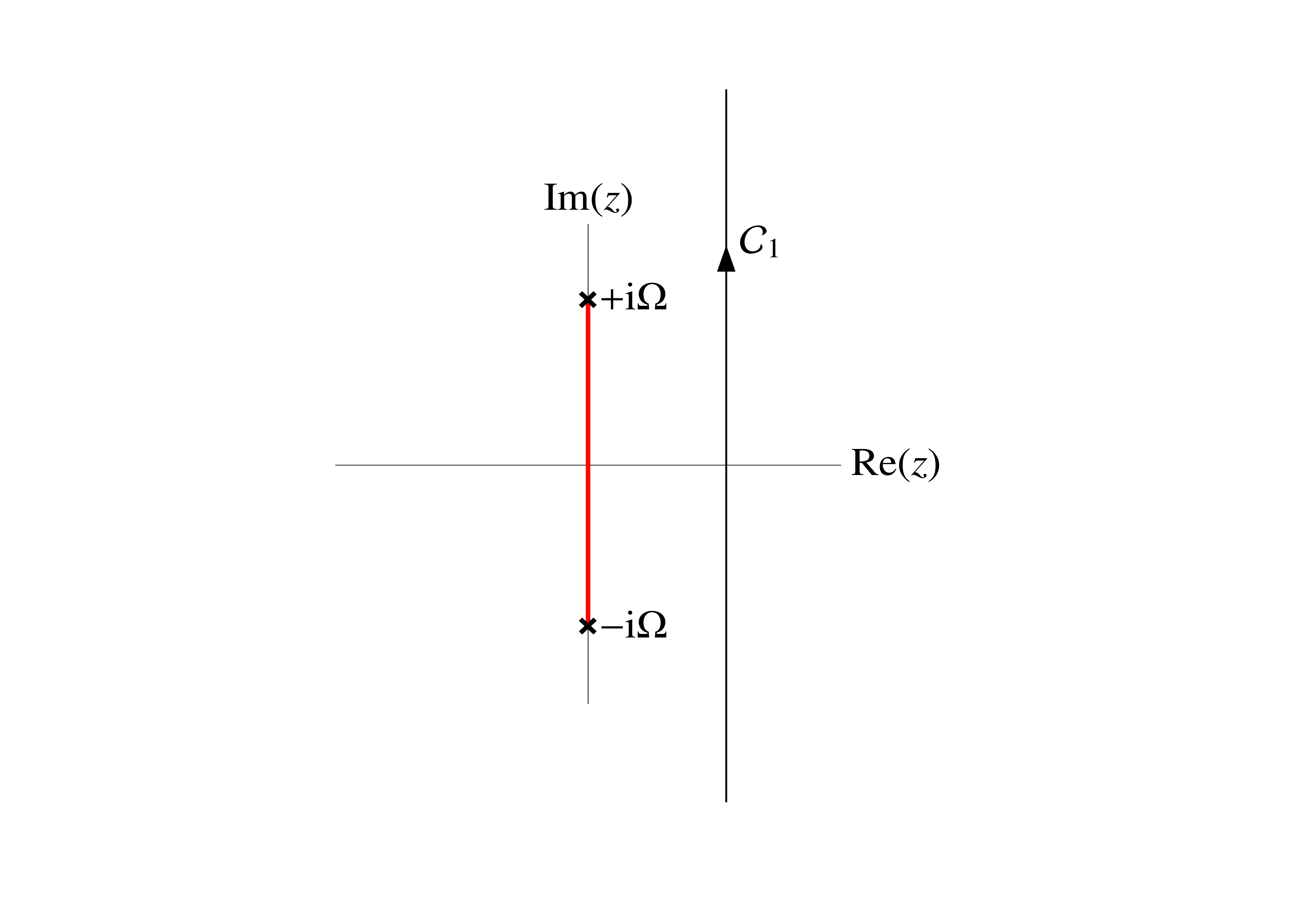}
\end{minipage}
\hspace{2mm}
\begin{minipage}{0.48\linewidth}
\vspace{1mm}\includegraphics[width=\linewidth, trim =25cm 0cm 25cm 0cm, clip]{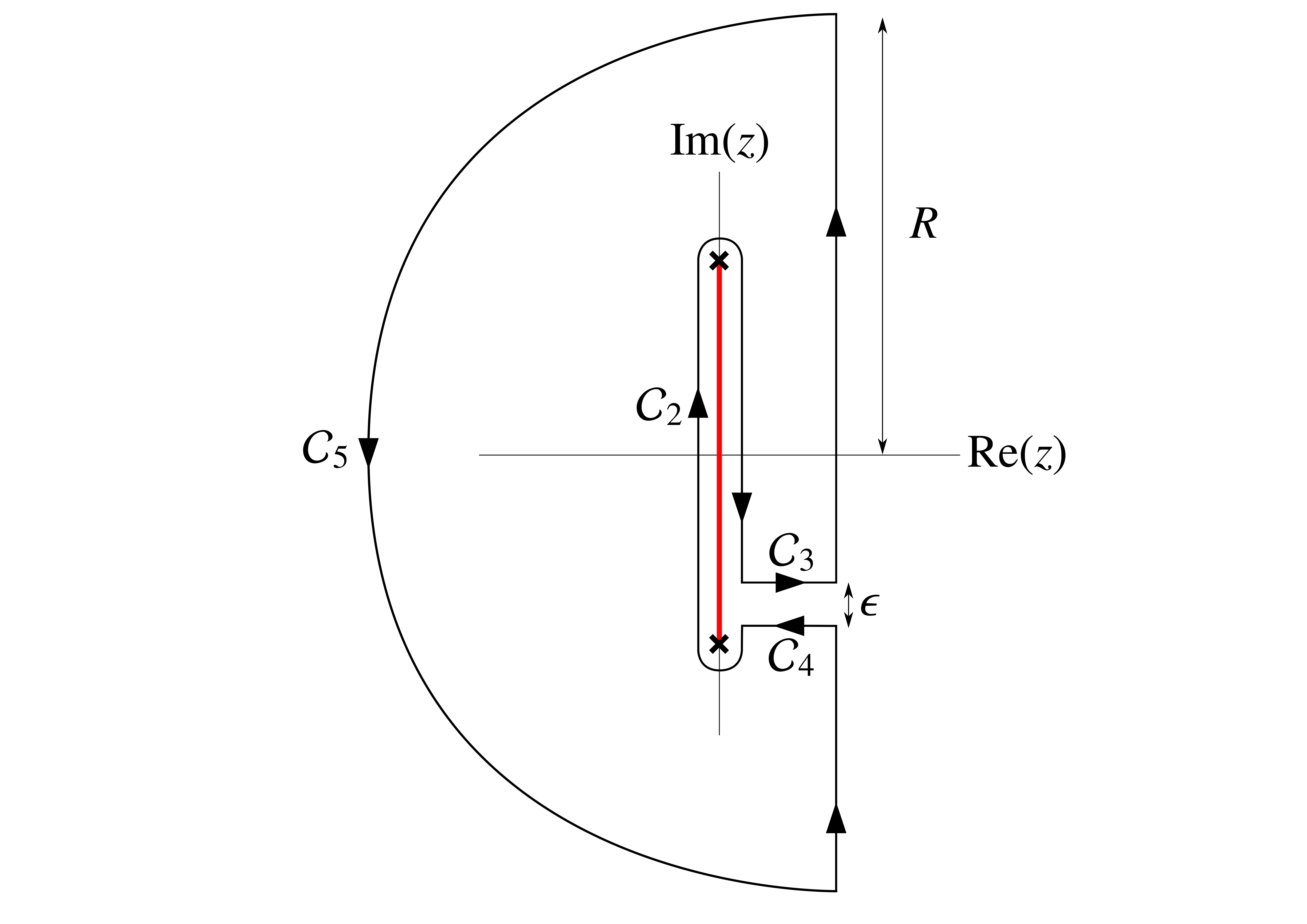}
\end{minipage}
\caption{Integration contours used to invert the Laplace transform $\tilde{G}_j(z)$, assuming that the inequality~\eqref{nopoleCondition} holds. (a)~A Bromwich contour lies to the right of the branch cut (thick red line) and extends parallel to the entire imaginary axis. (b)~A closed integration contour that avoids enclosing any singularities. Taking the limits $R\to \infty$ and $\epsilon\to 0$, the integrals along $\mathcal{C}_3$, $\mathcal{C}_4$ and $\mathcal{C}_5$ sum to zero, leaving only the contributions from $\mathcal{C}_1$ and $\mathcal{C}_2$.
\label{fig_contours}}
\end{figure}

The propagator is now obtained by carrying out the inverse Laplace transform, for $t>0$,
\begin{equation}
\label{Bromwich}
G_j(t) = \frac{1}{2\pi \ii}\int_{\mathcal{C}_1}\dd z\, \ee^{zt} \tilde{G}_j(z) = -\frac{1}{2\pi \ii}\int_{\mathcal{C}_2} \dd z\,\ee^{zt} \tilde{G}_j(z).
\end{equation}
Here, $\mathcal{C}_1$ denotes a Bromwich contour, while $\mathcal{C}_2$ is a closed contour encircling the branch cut in the clockwise sense. The two integration contours can be shown to give equal and opposite contributions using Cauchy's theorem (see Fig.~\ref{fig_contours}). After a change of variables to $\omega=\ii z$, the integral along $\mathcal{C}_2$ takes the form~\eqref{GjExact}, with $\varphi_j(\omega)$ given by Eq.~\eqref{varphi}.

It remains to justify Eqs.~\eqref{CinfExact} and \eqref{varThetaValue}. By taking the Laplace transform of Eq.~\eqref{GjExact}, we obtain
\begin{align}
\label{phiLaplace}
\tilde{G}_j(z) & = \int\dd\omega\, \frac{\varphi_j(\omega)}{z+\ii\omega}.
\end{align}
We therefore deduce that, in general,
\begin{align}
\label{LaplaceFourierConnection}
\lim_{\epsilon\searrow 0} \tilde{G}_j(\epsilon-\ii\omega) & = \int_0^\infty\dd t\, \ee^{-\ii\omega t} G_j(t)\notag\\
&  = \pi \varphi_j(\omega) +\ii \vartheta_j(\omega).
\end{align}
Using the first line, we obtain Eq.~\eqref{CinfExact} from Eq.~\eqref{CofTsolution}, while comparison of the second line with Eqs.~\eqref{GjOfZ} and \eqref{WofZexplicit} yields Eq.~\eqref{varThetaValue} for the uniform-chain environment model.

\section{Thermalisation at finite system-bath coupling}
\label{app_thermalisation}

In this appendix, we prove Eq.~\eqref{rhoInfThermalMarginal}. That is, when $\beta_\alpha = \beta$, $\mu_\alpha = \mu$ and $f_\alpha(\omega) = f(\omega)$, the open system equilibrates to a state $\r_{S\!}^\infty$ given by the reduction of a global Gibbs distribution. The following arguments can also be used to deduce Eq.~\eqref{rAlphaThermalMarginal} from Eq.~\eqref{rAlphaPop}.

Since the marginal of a Gaussian state is itself Gaussian~\cite{Cheong2004prb}, it suffices to check that Eq.~\eqref{rhoInfThermalMarginal} yields the correct correlation matrix~\eqref{Ceq}. Due to the symmetry $[\H,\hat{N}_j]=0$ of the global Hamiltonian, it follows that its equilibrium correlation matrix is diagonal, i.e.\ $\nolinebreak{\Tr[\ee^{-\beta(\H-\mu\hat{N})}\ddag_j\d_k]\propto\delta_{jk}}$. We explicitly compute the diagonal elements using a basis transformation to the eigenmodes of $\H$, given by~\cite{StefanucciVanLeeuwen} $\nolinebreak{\d_j = \sum_n \bra{0}\d_j\ket{n} \hat{e}_n}$, where $\nolinebreak{\ket{n} = \hat{e}^\dagger_n\ket{0}}$ is a single-particle eigenstate of $\H$ [cf.~Eq.~\eqref{phiSpectralFunction}] and the ladder operators $\hat{e}_n$ diagonalise the total Hamiltonian as $\nolinebreak{\H = \sum_n \varepsilon_n \hat{e}^\dagger_n\hat{e}_n}$. Substituting the above and using Eq.~\eqref{phiSpectralFunction}, straightforward manipulations lead to
\begin{equation}
\label{fullThermalCorrCalc}
\frac{\Tr\left [\ee^{-\beta\left (\H - \mu \hat{N}\right )} \ddag_j \d_j \right ]}{\ZZ\!\left (\beta, \H, \mu, \hat{N}\right )} =\int \dd\omega\, \varphi_j(\omega) f(\omega),
\end{equation}
in agreement with Eq.~\eqref{rhoInfThermalMarginal}, which completes the proof.

\section{Computing the currents}
\label{app_currents}

Due to the homogeneity of the steady-state currents, the particle current may be computed from the expectation value of either the intrasystem current \eqref{JpSystem} or the boundary current \eqref{JpBath}. The expected value of the intrasystem current is simply $\avg{\hat{J}^P_{S\!}}_t = g\Im[C_{12}(t)]$. In the limit $t\to \infty$, the particle current can thus be found directly from Eq.~\eqref{CinfExact}:
\begin{widetext}
\begin{equation}
\label{JpSysInf}
J^P_{S\!} = \frac{\pi g^2}{2\Delta}\int\dd\omega\,\mathcal{J}(\omega)\left [\varphi_1(\omega)\vartheta_2(\omega) - \varphi_2(\omega)\vartheta_1(\omega)\right ]\left [f_L(\omega) - f_R(\omega)\right ].
\end{equation}

To compute the expectation value of Eqs.~\eqref{JpBath} and \eqref{JeBath}, we make use of Eqs.~\eqref{dSolution} and \eqref{bSolution} and obtain, for example
\begin{align}
\label{JpBathAvg}
\avg{\hat{J}^P_{L}}_t & = 2\Im \left[ \mathsf{R}^\mathsf{T} \int\dd\omega\, \mathcal{J}(\omega) \mathsf{K}^\dagger(\omega,t) \mathsf{F}(\omega) \left ( \ee^{-\ii\omega t}\mathbbm{1} + \int\dd\omega'\, \mathcal{J}(\omega') \mathsf{I}(\omega',\omega,t)  \right )\mathsf{R} \right]_{11},\\
\label{JeBathAvg}
\avg{\hat{J}^E_{L}}_t & = 2\Im \left[ \mathsf{R}^\mathsf{T} \int\dd\omega\, \mathcal{J}(\omega) \mathsf{K}^\dagger(\omega,t) \mathsf{F}(\omega) \left ( \omega \ee^{-\ii\omega t} \mathbbm{1} + \int\dd\omega'\, \omega'\mathcal{J}(\omega') \mathsf{I}(\omega',\omega,t)  \right )\mathsf{R} \right]_{11},
\end{align}
where $\mathsf{K}(\omega,t) = {\rm diag}[K_1(\omega,t),K_2(\omega,t)]$ and $\mathsf{I}(\omega',\omega,t) = {\rm diag}[I_1(\omega',\omega,t),I_2(\omega',\omega,t)]$. In the limit $t\to \infty$, we find that
\begin{align}
\label{JpBathAvgLimit}
J^P_L & = 2\pi\sum_{k,l}R_{k1}R_{l1}\int\dd\omega\,\mathcal{J}(\omega) F_{kl}(\omega)\left \{\varphi_k(\omega) - \mathcal{J}(\omega)\left [\vartheta_k(\omega)\vartheta_l(\omega) + \pi^2 \varphi_k(\omega)\varphi_l(\omega)\right ]\right \},\\
J^E_L &= 2\pi\sum_{k,l}R_{k1}R_{l1}\int\dd\omega\,\omega\mathcal{J}(\omega) F_{kl}(\omega)\left \{\varphi_k(\omega) - \mathcal{J}(\omega)\left [\vartheta_k(\omega)\vartheta_l(\omega) + \pi^2 \varphi_k(\omega)\varphi_l(\omega)\right ]\right \}.
\end{align}
\end{widetext}
The above equations hold for arbitrary spectral densities. The explicit expressions~\eqref{JpInfExact}--\eqref{transmissionFunction} for the Newns spectral density~\eqref{Jnewns} can be derived by straightforward algebra with the help of Eq.~\eqref{varThetaValue}.

\section{Exponential-propagator approximation}
\label{app_epaPropagator}

\begin{figure}
\includegraphics[width=0.7\linewidth]{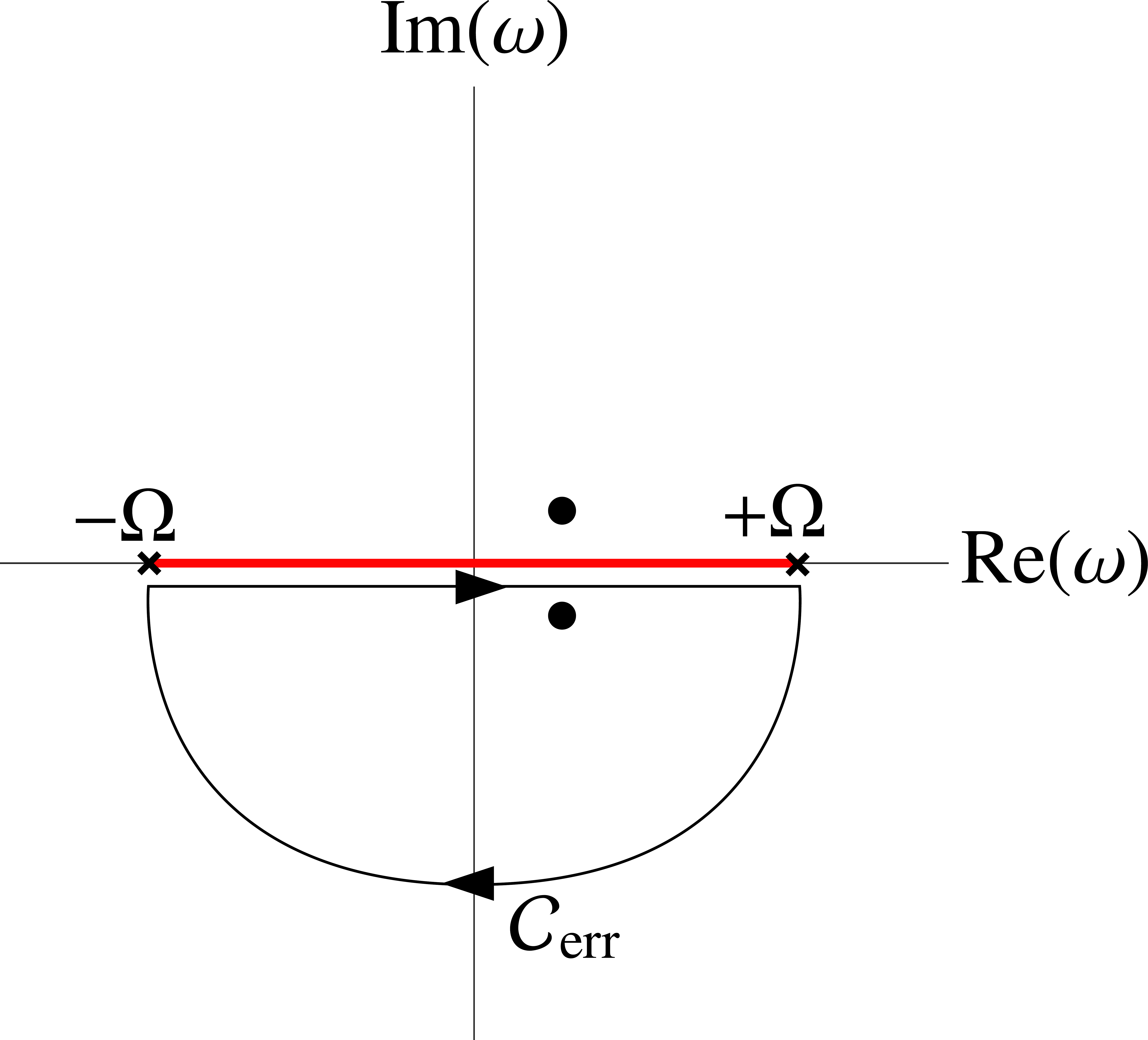}
\caption{Integration contour used to derive the EPA~\eqref{Gepa} when $\Gamma\ll \Omega$. Analytically continuing the integrand into the lower half of the complex $\omega$ plane yields the major contribution to the integral as the residue of the pole (solid circle) plus a small error contributed by the integral along the semi-circular arc $\mathcal{C}_{\rm err}$. 
\label{fig_epaContour}}
\end{figure}

In this appendix, we justify the exponential-propagator approximation~\eqref{Gepa} and prove that the error is of order $O(\Gamma/\Omega)$. The integral~\eqref{GjExact}, with $\varphi_j(\omega)$ given by Eq.~\eqref{varphi}, can be estimated with high accuracy for $\Gamma\ll \Omega$ by analytically continuing the integrand into the complex $\omega$ plane. Choosing (arbitrarily) the branch of the square root function whose real part is positive for $\Im \omega < 0$, we integrate along a closed contour encircling the pole at $\omega = E'_j - \ii\Gamma_j/2$ in the anti-clockwise sense (see Fig.~\ref{fig_epaContour}). This integral differs from the exact $G_j(t)$ by the contribution along the semi-circular arc $\mathcal{C}_{\rm err}$. We obtain
\begin{equation}
 \label{GexactSplit}
G_j(t) = p_j \ee^{-\ii E'_jt-\Gamma_j t} + {\rm err}_j(t).
 \end{equation} 
Here, $p_j$ is the residue at the pole,
\begin{equation}
\label{prefactor}
p_j = \sqrt{\frac{\Omega^2-(E'_j-\ii\Gamma_j/2)^2}{\Omega^2-\Gamma\Omega-E_j^2}},
\end{equation}
while the error term is given by
\begin{equation}
\label{errorDef}
{\rm err}_j(t) = - \int_{\mathcal{C}_{\rm err} } \dd \omega \; \ee^{-\ii\omega t}\varphi_j(\omega).
\end{equation}
A trivial rearrangement of terms leads to 
\begin{equation}
\label{Grearrange}
G_j(t) = \ee^{-\ii E'_jt-\Gamma_j t} + \left (p_j - 1\right )\ee^{-\ii E'_jt-\Gamma_j t}  + {\rm err}_j(t).
\end{equation}
The first term above is the EPA contribution~\eqref{Gepa}. The second term is a small relative correction to the leading-order result, which is clearly of first order in $\Gamma/\Omega$ and decays exponentially in time. Finally, the remaining absolute error $\lvert{\rm err}_j(t)\rvert$ can be bounded as 
\begin{align}
\label{errorBound}
\lvert{\rm err}_j(t)\rvert & \leq \frac{\Omega}{1-\Gamma/\Omega}\int_{0}^\pi \dd \theta \; \left \lvert \frac{\ee^{-\ii(\theta+ \Omega t\ee^{-\ii\theta})} \mathcal{J}_{\rm N}(\Omega\ee^{-\ii\theta})}{(\Omega\ee^{-\ii\theta} - E'_j)^2 + \Gamma_j^2/4}\right\rvert \notag \\
& = \frac{\Gamma\Omega^2}{\sqrt{2}\pi(\Omega-\Gamma)} \int_0^\pi \dd\theta\, \frac{\ee^{-\Omega t\sin\theta} \sin^{1/2}\theta}{\left \lvert(\Omega\ee^{-\ii\theta} - E'_j)^2 + \Gamma_j^2/4\right \rvert}\notag \\
& < \frac{\Gamma}{\Omega} \times\frac{1}{(1 -\Gamma/\Omega)}\frac{\Omega^2}{{(\Omega - E'_j)^2 + \Gamma_j^2/4}} u(\Omega t).
\end{align}
On the first line above, we changed variables to $\omega = \Omega \ee^{-\ii\theta}$ and used the fact that $\lvert\int\dd x\, f(x)\rvert \leq \int\dd x\,\lvert f(x)\rvert$, while the third line follows from setting the denominator of the integrand to its minimum value and defining the dimensionless function
\begin{align}
\label{uFunc}
u(\tau) & = \frac{1}{\sqrt{2}\pi}\int_0^\pi \dd\theta\,\ee^{-\tau\sin\theta} \sin^{1/2}\theta 	\notag \\
& = \frac{\sqrt{\pi \tau}}{2} \left(I_{-\frac{3}{4}}(\tau/2) I_{\frac{1}{4}}(\tau/2)-I_{-\frac{1}{4}}(\tau/2)I_{\frac{3}{4}}(\tau/2)\right),
\end{align}
where $I_s(x)$ is a modified Bessel function of the first kind. The function $u(\tau)$ is positive, monotonically decreasing and vanishes as $\tau\to \infty$ with the limiting behaviour $\lim_{\tau\to\infty}\tau^{3/2}u(\tau)=(2\pi)^{-1/2}$. Hence, by replacing $u(\Omega t)$ in Eq.~\eqref{errorBound} by $u(0)\approx 0.54$ one obtains a time-independent upper bound on $\lvert{\rm err}_j(t)\rvert$ that is of order $O(\Gamma/\Omega)$, as claimed. Note, however, that the magnitude of the error also depends on $E_j/\Omega$. In particular, the error is larger when $E_j$ is closer to the band edge at $\pm \Omega$. 

\section{Entropy-production inequality}
\label{app_entropyProduction}

This appendix demonstrates the entropy-production inequality~\eqref{secondLawLimit}, starting from the Spohn inequality~\eqref{Spohn}. We first demonstrate that $\LL_{\rm int}$ does not generate a positive evolution, i.e.\ the map $\ee^{\LL_{\rm int}t}$ is not positive. We write $\LL_{\rm int}$ in diagonal form
\begin{equation}
\label{genDiag}
\LL_{\rm int} = \sum_{j=1}^4 \lambda_j \DD[\hat{L}_j],
\end{equation}
where $\lambda_1 = -\lambda_2 = -\lambda_3 = \lambda_4 = \lvert \Lambda^+_{12}\rvert$, $\hat{L}_1 = \hat{L}_2^\dagger = \tfrac{1}{\sqrt{2}}(\d_1 - \ee^{-\ii\theta}\d_2)$, $\hat{L}_3= \hat{L}_4^\dagger = \tfrac{1}{\sqrt{2}}(\d_1+\ee^{-\ii\theta}\d_2)$, and $\theta = \arg(\Lambda^+_{12})$. Note that these Lindblad operators satisfy the canonical anti-commutation relations $\{ \hat{L}_j,\hat{L}_k^\dagger\} = \delta_{jk}$, and are thus linearly independent, i.e. $\nolinebreak{\Tr[\hat{L}_j^\dagger \hat{L}_k ] = 2\delta_{jk}}$. Therefore, $\LL_{\rm int}$ generates a positive evolution if and only if
\begin{equation}
\label{positiveCondition}
\sum_{j=1}^4 \lambda_j \left \lvert\matrixel{a}{\hat{L}_j}{b}\right \rvert^2\geq 0,
\end{equation}
for every orthonormal pair of states $\ket{a}$ and $\ket{b}$~\cite{Breuer2016rmp}. However, it is straightforward to find a counterexample. Consider, for instance, the choice $\ket{b} = \hat{L}^\dagger_3\ket{0}$ and $\ket{a} = \ket{0}$. Then $\bra{a}\hat{L}_j\ket{b} = \delta_{j3}$ and, since $\lambda_3 < 0$, inequality~\eqref{positiveCondition} does not hold. Hence, the Spohn inequality cannot be directly applied to $\LL_{\rm int}$~\cite{MullerHermes2017ahp}.

Nevertheless, the Lindblad generators $\LL_\alpha$ do satisfy the Spohn inequality~\eqref{Spohn}. Hence, taking $\r_{S\!}(t) = \r^\infty_{S\!}$ and summing over the baths, we obtain
\begin{equation}
\label{secondLawFiniteGamma}
\Tr\left [\ln\r^\infty_{S\!}\LL_{\rm int}\r_{S\!}^\infty\right ] -\sum_{\alpha = L,R} \beta_\alpha \avg{\LL^\dagger_\alpha\left (\H_{S\!,\alpha}^* - \mu_\alpha \hat{N}_{S\!} \right )}_\infty  \geq 0.
\end{equation}
To obtain the first term on the left-hand side (LHS), we have used the stationary property $\nolinebreak{\sum_\alpha \LL_\alpha \r^\infty_{S\!} = -\LL_{\rm int}\r^\infty_{S\!}}$. In the second term, we introduced the quantum Hamiltonian of mean force~\cite{Campisi2009prl}
\begin{equation}
\label{meanForce}
\H^*_{S,\alpha} = -T_\alpha\ln \left \lbrace\frac{\Tr_B\left [\ee^{-\beta_\alpha\left (\H-\mu_\alpha \hat{N}\right )} \right ]  }{\ZZ\!\left (\beta_\alpha,\H_B,\mu_\alpha,\hat{N}_B\right )}\right \rbrace+\mu_\alpha\hat{N}_{S\!},
\end{equation}
which is defined so that [cf. Eq.~\eqref{rAlphaThermalMarginal}]
\begin{equation}
\label{meanForceRalpha}
\hat{r}_\alpha = \frac{\ee^{-\beta_\alpha \left (\H^*_{S\!,\alpha}-\mu_\alpha \hat{N}_{S\!}\right )}}{\ZZ\!\left (\beta_\alpha,\H^*_{S\!,\alpha},\mu_\alpha,\hat{N}_{S\!}\right )}.
\end{equation}

\begin{figure}
\includegraphics[width=0.8\linewidth]{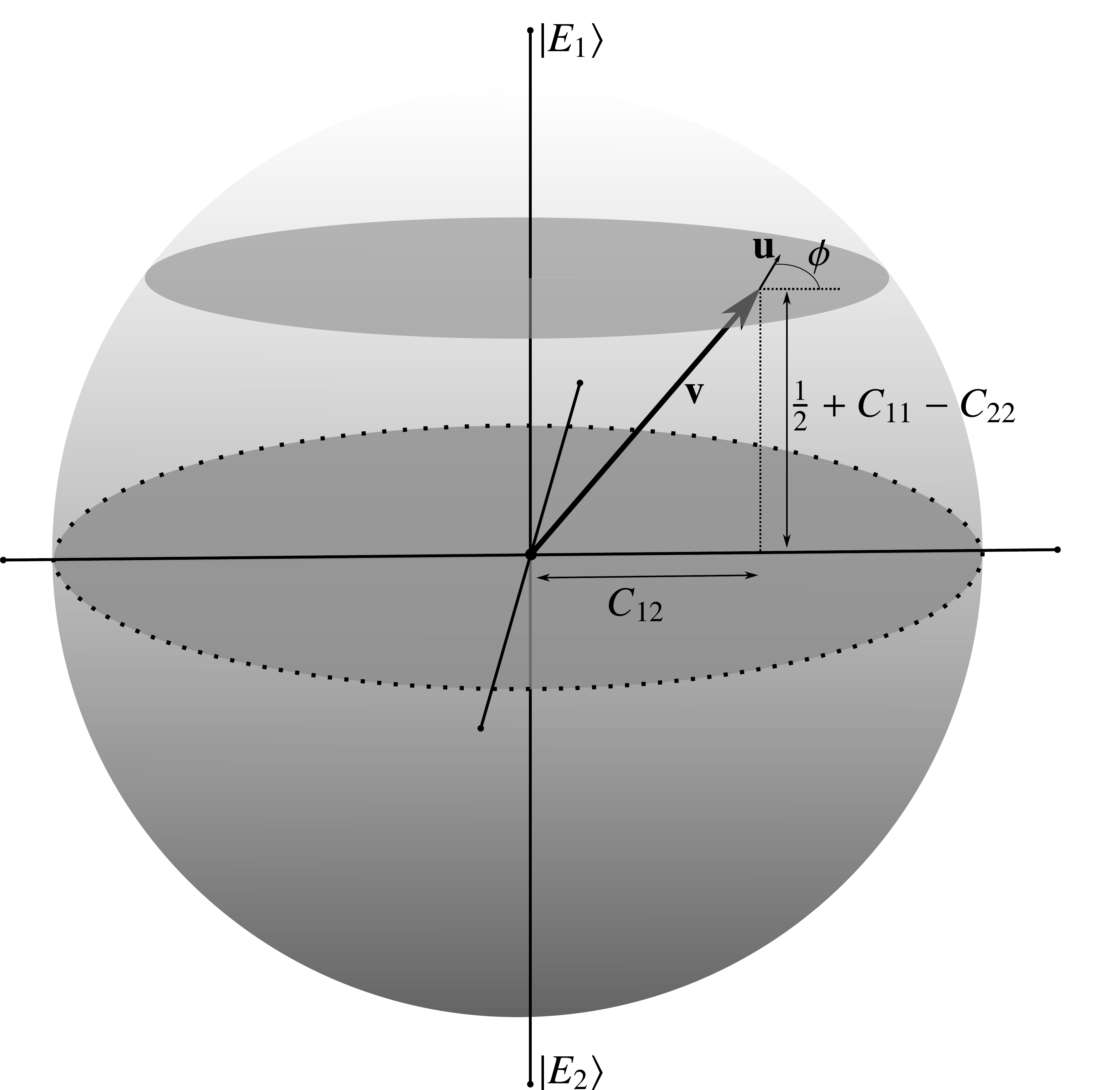}
\caption{Depiction of the NESS restricted to the single-particle subspace. The state, described by a Bloch vector $\mathbf{v}$, is displaced by a flow $\mathbf{u}$ generated by $\LL_{\rm int}$, which always points towards the surface of the Bloch sphere and thus decreases the entropy.\label{fig_BlochSphereEntropy}}
\end{figure}

Let us examine the first term on the LHS of inequality~\eqref{secondLawFiniteGamma} more closely. This term represents the negative time derivative of the system's von Neumann entropy generated by $\LL_{\rm int}$. This term is non-negative, which can be seen by the following geometrical argument, illustrated in Fig.~\ref{fig_BlochSphereEntropy}. According to Eq.~\eqref{LintPopCoh}, $\LL_{\rm int}$ changes only the coherences. Hence, it suffices to consider the action of $\LL_{\rm int}$ in the two-dimensional subspace spanned by the single-particle states $\lvert E_j\rangle = \ddag_j\ket{0}$. The restriction of the density matrix $\r^\infty_{S\!}$ to the single-particle subspace is represented by a Bloch vector, $\mathbf{v}$, which we choose to lie perpendicular to the $y$ axis. Specifically, the projection of $\mathbf{v}$ along the $z$ axis is given by $\tfrac{1}{2}+ C_{11}-C_{22}$, while the projection along the $x$ axis is $C_{12}$, where all quantities are evaluated in the limit $t\to \infty$. 

Consider now the flow generated by $\LL_{\rm int}$ over a small time interval $\delta t$, i.e.\ the vector $\mathbf{u}$ such that the state $\ee^{\LL_{\rm int}\delta t}\r^\infty_{S\!}$ corresponds to the shifted Bloch vector $\mathbf{v} +\mathbf{u}\delta t+ O(\delta t^2)$. According to Eq.~\eqref{LintPopCoh}, $\mathbf{u}$ lies in the plane perpendicular to the $z$ axis and points in a direction determined by the complex argument of $\Lambda^+_{12}$. Using Eq.~\eqref{Ceom}, we deduce that $\nolinebreak{\Lambda_{12}^+=(\Tr[\mathsf{\Gamma}]/2 -\ii\Delta)C_{12}}$. Thus, the angle subtended by $\mathbf{u}$ from the $x$ axis is $\nolinebreak{\phi = \arctan(2\Delta/\Tr[\mathsf{\Gamma}])}$, such that $\nolinebreak{0 \leq \phi \leq \pi/2}$. It follows that the flow generated by $\LL_{\rm int}$ never moves the Bloch vector away the surface of the Bloch sphere, i.e.\ the length of $\mathbf{v}$ does not decrease. Since the von Neumann entropy is a monotonically decreasing function of the length of $\mathbf{v}$, we conclude that it is a non-increasing function under this flow. 

Now let us demonstrate that the derivative of the von Neumann entropy generated by $\LL_{\rm int}$ behaves as $o(\Gamma)$ in the limit $\Gamma\to 0$. For this it is convenient to use an explicit representation of the Gaussian NESS,
\begin{equation}
\label{rhoSgaussian}
\r^\infty_{S\!} = \frac{\exp\left (- \mathbf{\ddag}\mathsf{P}^{\mathsf{T}}\mathbf{\d}\right )}{\det\left [1+ \ee^{-\mathsf{P}}\right ]},
\end{equation}
where $\mathsf{P}$ is a positive semi-definite matrix that satisfies $\nolinebreak{\mathsf{C} = [\ee^{\mathsf{P}}+ 1]^{-1}}$. Then, making use of Eq.~\eqref{LintPopCoh}, we find the explicit representation
\begin{equation}
\label{LambdaPentropy}
\Tr\left [\ln\r^\infty_{S\!}\LL_{\rm int}\r_{S\!}^\infty\right ] = -2\Re \left [\Lambda_{12}^+ P_{21}\right ].
\end{equation}
Clearly, $\Lambda_{12}^+ = O(\Gamma)$ by its definition~\eqref{LambdaPlusInf}. Furthermore, since the coherence in the stationary state $C_{12} = \Lambda_{12}^+/[\Tr[\mathsf{\Gamma}]/2 -\ii\Delta]$ also vanishes as $\Gamma\to 0$, we conclude that $P_{21}\to 0$ as $\Gamma\to 0$. Overall, the contribution from the expression~\eqref{LambdaPentropy} therefore vanishes upon dividing the relation~\eqref{secondLawFiniteGamma} by $\Gamma$ and taking the limit $\Gamma\to 0$. Taking into account the fact that $\lim_{\Gamma\to 0}\H^*_{S\!,\alpha} = \H_{S\!}$, the inequality~\eqref{secondLawLimit} is thus recovered.

\clearpage

\end{document}

%% file: packages.tex
\usepackage[T1]{fontenc}
\usepackage{graphicx,epsfig}
\usepackage{amsmath,amssymb,amsthm,bbm,bm}
\usepackage{fancyhdr}
\usepackage{txfonts}
\usepackage{libertine}
\usepackage{import}
\usepackage{booktabs}
\usepackage{exscale}
\usepackage[english]{babel}
\usepackage{color}
\usepackage{appendix}
\usepackage{dsfont}
\usepackage{hyperref}
\usepackage{xcolor}
\hypersetup{
    colorlinks,
    linkcolor={rgb:red,2; blue,1},
    citecolor={blue!60!black},
    urlcolor={blue!60!black}
}
\usepackage{footmisc}

%% file: definitions.tex
\renewcommand{\d}{\dagger}

\renewcommand{\H}{\hat{H}}

\newcommand{\ii}{\ensuremath{\mathrm{i}}}
\newcommand{\ee}{\ensuremath{\mathrm{e}}}
\newcommand{\dd}{\mathrm{d}}

\newcommand{\bra}[1]{\ensuremath{\left\langle #1 \right\rvert}}
\newcommand{\ket}[1]{\ensuremath{\left\lvert #1 \right\rangle}}

\newcommand{\matrixel}[3]{\ensuremath{\left\langle #1 \vphantom{#2#3} \middle\vert #2 \middle| #3 \vphantom{#1#2} \right\rangle}}

\newcommand{\avg}[1]{\left \langle #1 \right \rangle}

\renewcommand{\Re}{\,\mathrm{Re}\,}
\renewcommand{\Im}{\,\mathrm{Im}\,}

\newcommand{\Tr}{\mathrm{Tr}}

\renewcommand{\a}{\ensuremath{\hat{a}}}
\newcommand{\adag}{\ensuremath{\hat{a}^\dagger}}
\renewcommand{\b}{\ensuremath{\hat{b}}}
\newcommand{\bdag}{\ensuremath{\hat{b}^\dagger}}
\renewcommand{\c}{\ensuremath{\hat{c}}}
\newcommand{\cdag}{\ensuremath{\hat{c}^\dagger}}
\renewcommand{\d}{\ensuremath{\hat{d}}}
\renewcommand{\ddag}{\ensuremath{\hat{d}^\dagger}}

\newcommand{\td}{\ensuremath{\tilde{d}}}

\renewcommand{\H}{\ensuremath{\hat{H}}}
\renewcommand{\r}{\ensuremath{\hat{\rho}}}

\newcommand{\LL}{\ensuremath{\mathcal{L}}}

\newcommand{\PP}{\ensuremath{\mathcal{P}}}

\newcommand{\DD}{\ensuremath{\mathcal{D}}}

\newcommand{\ZZ}{\ensuremath{\mathcal{Z}}}

\newcommand{\SJ}{\mathcal{J}}

\newcommand{\eqr}[1]{Eq.~\eqref{#1}}